\documentclass{jfm}
\usepackage{amsmath}
\usepackage{siunitx}
\usepackage{natbib}
\DeclareMathOperator{\erf}{erf}
\graphicspath{{./pictures/}}

\title{Entrainment, diffusion and effective compressibility in a self-similar turbulent jet}

\author{Thomas Basset\aff{1} \corresp{\email{thomas.basset@ens-lyon.fr}}, Bianca Viggiano\aff{2}, Thomas Barois\aff{3}, Mathieu Gibert\aff{4}, Nicolas Mordant\aff{5}, Ra\'ul Bayo\'an Cal\aff{2}, Romain Volk\aff{1} \and Micka\"el Bourgoin\aff{1}}
\affiliation{\aff{1}Univ Lyon, ENS de Lyon, CNRS, Laboratoire de Physique, F-69342 Lyon, France
\aff{2}Department of Mechanical and Materials Engineering, Portland State University, Portland, OR 97201, USA
\aff{3}Univ. Bordeaux, CNRS, LOMA, UMR 5798, F-33400 Talence, France
\aff{4}Univ. Grenoble Alpes, CNRS, Grenoble INP, Institut N\'eel, 38000 Grenoble, France
\aff{5}Univ. Grenoble Alpes, CNRS, Grenoble INP, LEGI, 38000 Grenoble, France}
\shortauthor{T. Basset and others}

\begin{document}

\maketitle

\begin{abstract}
An experimental Lagrangian study based on particle tracking velocimetry has been completed in an incompressible turbulent round water jet freely spreading into water. The jet is seeded with tracers only through the nozzle: inhomogeneous seeding called \textit{nozzle seeding}. The Lagrangian flow tagged by these tracers therefore does not contain any contribution from particles entrained into the jet from the quiescent surrounding fluid. The mean velocity field of the nozzle seeded flow, $\langle \boldsymbol{U_\varphi} \rangle$, is found to be essentially indistinguishable from the global mean velocity field of the jet, $\langle \boldsymbol{U} \rangle$, for the axial velocity while significant deviations are found for the radial velocity. This results in an effective compressibility of the nozzle seeded flow for which $\bnabla \bcdot \langle \boldsymbol{U_\varphi} \rangle \neq 0$ even though the global background flow is fully incompressible. By using mass conservation and self-similarity, we quantitatively explain the modified radial velocity profile and analytically express the missing contribution associated to entrained fluid particles. By considering a classical advection-diffusion description, we explicitly connect turbulent diffusion of mass (through the turbulent diffusivity $K_T$) and momentum (through the turbulent viscosity $\nu_T$) to entrainment. This results in new practical relations to experimentally determine the non-uniform spatial profiles of $K_T$ and $\nu_T$ (and hence of the turbulent Prandtl number $\sigma_T = \nu_T/K_T$) from simple measurements of the mean tracer concentration and axial velocity profiles. Overall, the proposed approach based on nozzle seeded flow gives new experimental and theoretical elements for a better comprehension of turbulent diffusion and entrainment in turbulent jets.
\end{abstract}


\section{Introduction}
\begin{figure}
(\textit{a}) \hspace{0.4\textwidth}(\textit{b})\\
\centerline{\includegraphics[width=5.15cm,height=10cm]{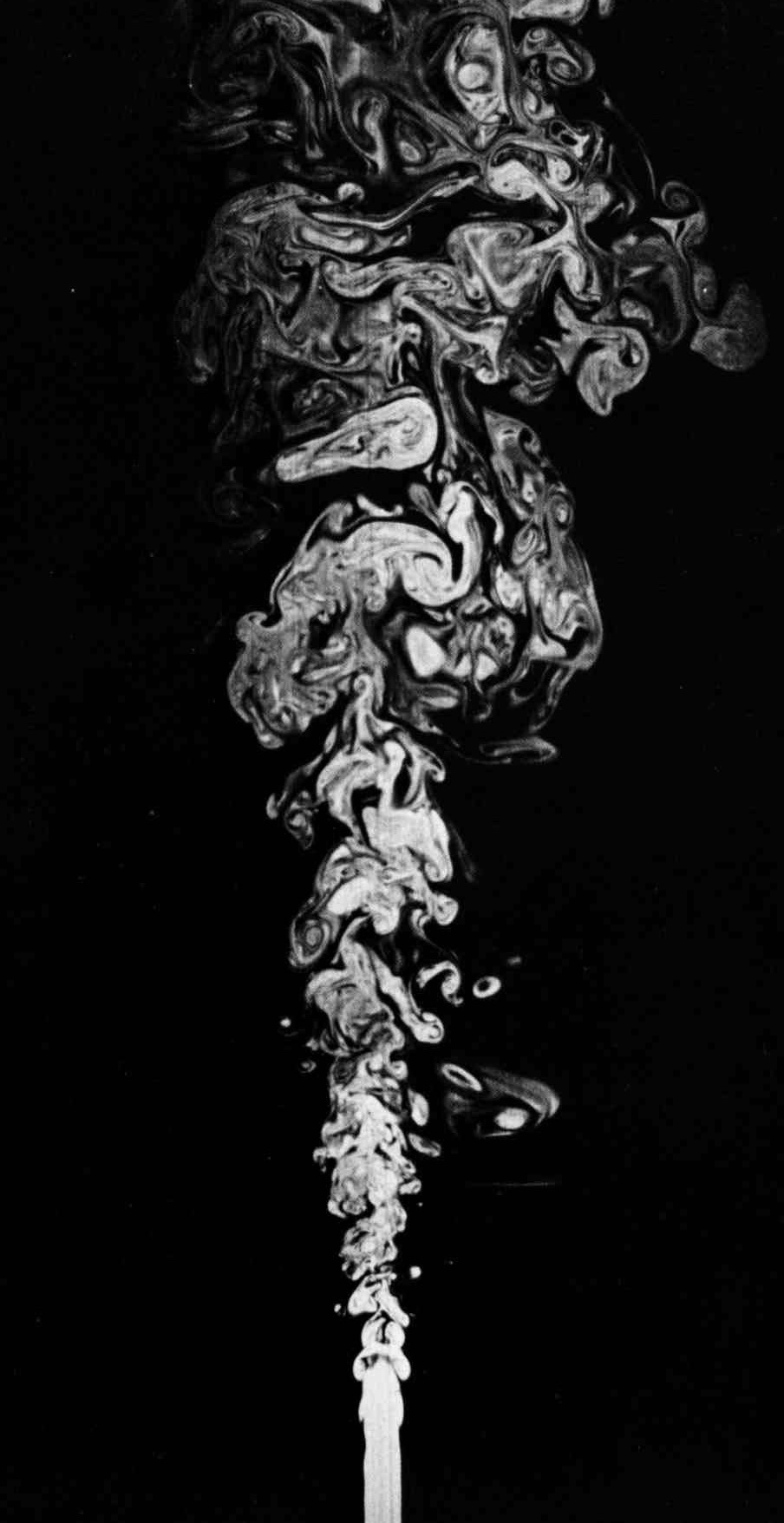}
\includegraphics[height=10cm]{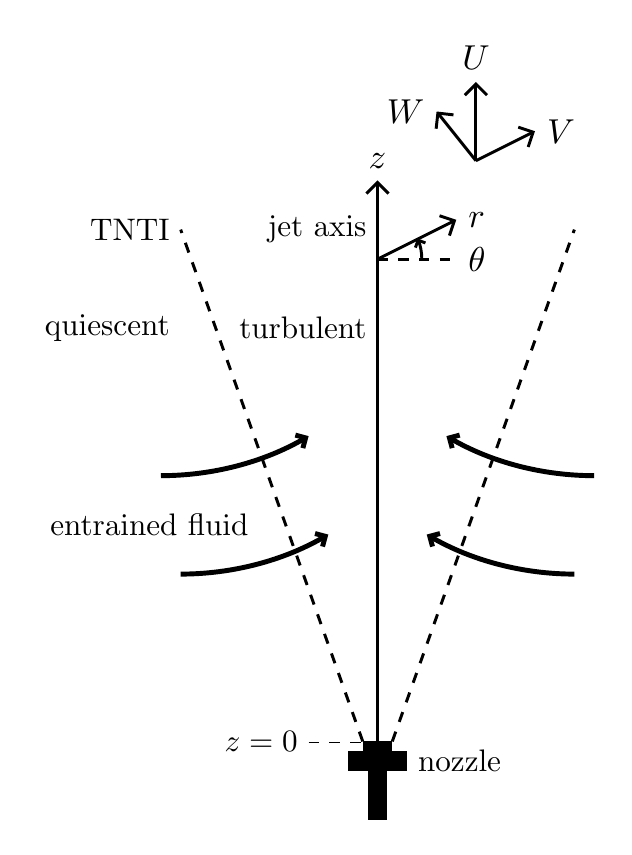}}
\caption{(\textit{a})~Laser-induced fluorescence of a turbulent round water jet spreading into water (adapted from \citet{vandyke1982album}, based on \citet{dimotakis1983structure}). Fluorescent dye is injected through the nozzle, thus white fluid comes from the nozzle and black fluid from the ambient. We can observe that initially quiescent fluid is entrained up to the turbulent core of the jet. (\textit{b})~Schematic of the jet with cylindrical coordinates $(z,r,\theta)$ and velocity components $U$, $V$ and $W$ (two-dimensional projection of a three-dimensional jet). The turbulent core of the jet is fed with entrained fluid.\label{fig:jet}}
\end{figure}
Free shear flows, such as jets, wakes or mixing layers, are common flows in nature, industry and laboratory, with turbulence arising from mean velocity differences, i.e. from shearing \citep{pope2000turbulent}. The incompressible free round jet, which is the flow studied in this article, is a simple configuration generated by a high-speed fluid issuing from a small source (\textit{nozzle}) into a large reservoir with quiescent fluid. The jet eventually grows into a flow which is statistically stationary, though inhomogeneous in space, with a turbulent core surrounded by a slow (almost at rest) non-turbulent flow. Parcels of fluid from the quiescent region are constantly crossing the turbulent/non-turbulent interface (TNTI) feeding the jet \citep{zhou2020energy, cafiero2020nonequilibrium}, a process called \textit{entrainment} \citep{corrsin1955free, philip2012large}. The overall dynamics within the core of the jet therefore results from both contributions: fluid parcels which have been injected through the nozzle together with fluid parcels which have been entrained from the ambient. It can be observed in figure~\ref{fig:jet}(\textit{a}) where fluid coming from the nozzle and fluid from the ambient are highly mixed. Figure~\ref{fig:jet}(\textit{b}) presents a schematic of the jet and entrainment process with the notations used in the following.

The major relevance for many natural and industrial systems (volcanic eruptions, sprays, rocket exhaust, chemical injectors, etc.) together with remarkable properties of free round jets have motivated numerous theoretical and experimental studies of this flow over almost a century \citep{corrsin1943investigation, hinze1949transfer, corrsin1950further, wygnanski1969measurements, panchapakesan1993turbulence1, hussein1994velocity, pope2000turbulent, schlichting2017boundary}. One of the most remarkable properties revealed by these studies is that, sufficiently far downstream from the nozzle (typically a few tens of nozzle diameters $D$), free round jets become self-similar with increasing downstream distance $z$ from the nozzle: the spatial dependence of velocity statistics (including the mean and fluctuating axial and radial velocity profiles) can be simply rescaled and expressed in terms of a single spatial variable $\eta = r/z$, where $r$ is the radial coordinate (note that due to axisymmetry, the statistics of free round jets are trivially independent of the circumferential coordinate $\theta$). Interestingly, self-similarity does not only hold for the kinematic properties of the jet, but also for its mixing properties. For instance, if a passive scalar (temperature, dye, aerosol, etc.) is injected through the nozzle, the streamwise evolution of the concentration field also exhibits self-similarity with spatial profiles only dependent on the self-similar variable $\eta = r/z$ \citep{dowling1990similarity}.

Self-similarity has profound consequences, both on physical properties and on the development of reduced models for the jet. From the physical point of view, one of the most celebrated consequences of self-similarity is for example that in a free round jet the turbulent Reynolds number $\Rey$ in the self-similar region is independent of the distance to the nozzle \citep{pope2000turbulent}. On the modelling side, self-similarity combined to other relevant approximations (such as the turbulent boundary-layer equations) allows derivation of analytical solutions for the jet velocity and concentration profiles, in terms of effective turbulent transport coefficients such as the turbulent viscosity $\nu_T$ and the turbulent diffusivity $K_T$ (related by the turbulent Prandtl number $\sigma_T = \nu_T/K_T$). These coefficients are crucial to model the turbulent mixing of passive scalars injected through the nozzle \citep{batchelor1957diffusion, chua1990turbulent, tong1995passive, pope2000turbulent, chang2002turbulent}. However, in spite of the relatively deep knowledge achieved today on free round jets, important questions still remain, even regarding such simple large-scale momentum and mass transport properties. In particular, the precise role of entrainment on the self-similar velocity and concentration profiles, on the momentum and mass transport coefficients and on their eventual spatial inhomogeneity is not yet elucidated.

Numerous studies have been realised to characterise entrainment, from simulations \citep{mathew2002characteristics, watanabe2016lagrangian} to particle image velocimetry \citep{westerweel2005mechanics, westerweel2009momentum, mistry2016entrainment, mistry2019kinematics} and particle tracking velocimetry \citep{wolf2012investigations}. Nevertheless they have mainly focused on the dynamics of the TNTI and the mechanisms in its vicinity by which ambient parcels of fluid get trapped into the core of the jet, generally distinguishing the role of large-scale structures (\textit{engulfment}) and small-scale eddy motions (\textit{nibbling}) \citep{philip2012large}. We do not address here such, rather local, entrainment mechanisms, but rather question, in a Lagrangian perspective (entrainment is innately Lagrangian), the impact of entrainment on the global Eulerian properties of the turbulent core of the jet. In other words, when describing the large-scale characteristics of the jet, such as the self-similar mean axial and radial velocity profiles and the turbulent viscosity and diffusivity, can we distinguish (and eventually separate) the contribution from fluid parcels which have been injected through the nozzle (which we shall call in the sequel \textit{nozzle seeded particles}) and that from fluid parcels which have been entrained into the jet (which we shall call in the sequel \textit{entrained particles})? The question is far from rhetorical as in many practical situations nozzle seeded and entrained particles are physically distinct, though coupled. It is the case for instance of sprays, eruptions, chimneys, etc., where actual particles or parcels of fluid carrying a passive scalar (concentration field, temperature, etc.) of interest are injected solely through the nozzle although their subsequent spread is affected by their coupling with the parcels of fluid entrained from the ambient medium. How deep into the core of the jet do entrained particles influence the dynamics of nozzle seeded particles? How substantial is their influence on the effective transport coefficients? In particular, can we quantitatively measure and/or predict the influence of entrained particles on the dispersion of nozzle seeded particles? Is this influence homogeneous in space or does it impact differently the borders and the centre of the jet? Such are the questions we aim to address in the present article.

In reference Eulerian measurements (such as hot-wire anemometry) carried out to characterise turbulence in jets, both contributions are naturally entangled as the sensor does not distinguish the origin (nozzle or ambient) of the fluid parcels it is probing. The distinction between nozzle seeded and entrained particles is intrinsically Lagrangian as it concerns specifically tagged particles according to the initial position of their trajectories. With this respect, this distinction can also be investigated with Eulerian measurement techniques based on particles, such as particle image velocimetry or laser Doppler velocimetry, if they are used with the Lagrangian conditioning presented in the next paragraph. This inhomogeneous seeding situation differs from the usual homogeneous seeding required to access truly Eulerian fields. Beyond the fundamental aspect of disentangling the role of nozzle seeded and entrained particles on the overall jet dynamics, this distinction is also of relevance for applications such as particle-laden jets and the mixing of a passive scalar injected within the jet. In such situations, particles (or substances) come from the nozzle and get dispersed as they mix with entrained particles. Note that in particle-laden jets, the dynamics of the particles may be further complicated by their finite inertia (related to their finite size and/or density mismatch relative to the carrier flow). We do not address in the present work the role of inertia, and will only consider the case of Lagrangian tracers whose dynamics reflects that of fluid parcels.

To achieve such a Lagrangian distinction, the present study focuses on the dynamics of tracer particles solely injected through the nozzle of the jet (\textit{nozzle seeding}), which we compare to the known behaviour for the global Eulerian properties of the jet, which naturally includes both (nozzle seeded and entrained) contributions. Our study combines experimental measurements together with new theoretical formulations derived specifically for the sole contribution of the flow tagged by nozzle seeded particles, and accounting for mass conservation and self-similarity. By doing so, several remarkable findings are obtained:

\begin{itemize}
\item First, we experimentally show that the mean axial velocity profile associated to nozzle seeded particles marginally differs from the global Eulerian profile. Whereas the measured radial velocity profile is found compressible: the continuity equation, ensuring the zero-divergence of the global Eulerian velocity field, is only fulfilled if both (nozzle seeded and entrained) are considered together and not separately.
\item Second, this observation leads to the consideration of the tracer concentration field for the continuity equation. A simple relation between the axial and the radial mean velocity profiles of the nozzle seeded flow is found and, by comparison to its well-known counterpart for the global Eulerian description of the jet, allows clear identification of the contribution due to entrainment up to the core of the jet.
\item Third, by describing the dispersion of nozzle seeded particles as a classical advection-diffusion process, we relate the turbulent diffusivity $K_T(\eta)$ (which is assumed space-dependent and self-similar) to the effective compressibility of the nozzle seeded flow previously mentioned, hence to the entrainment process. Based on this relation, we propose a novel approach to measure the spatial profile of $K_T(\eta)$, which is found to depend on the mean axial velocity and tracer concentration profiles. This approach can be extended to the estimate of the turbulent viscosity $\nu_T(\eta)$, which follows a similar relation and thus is also related to entrainment. Finally, combining these two quantities, we derive a simple expression of the turbulent Prandtl number $\sigma_T(\eta)$ which is experimentally measured.
\end{itemize}

In §~\ref{sec:exp}, we present the experimental set-up and particle tracking methods used to characterise the dynamics of nozzle seeded particles. Sections~\ref{sec:velocity} and~\ref{sec:velocity_phi} provide experimental and theoretical results for the mean axial and radial velocities of the flow associated to nozzle seeded particles. In §~\ref{sec:diff}, results about turbulent transport coefficients based on advection-diffusion model are reported. Finally, main conclusions are summarised in §~\ref{sec:conclusion}.

\section{Experimental methods}\label{sec:exp}
\subsection{Experimental set-up}
\begin{figure}
\centerline{\includegraphics[width=0.7\textwidth]{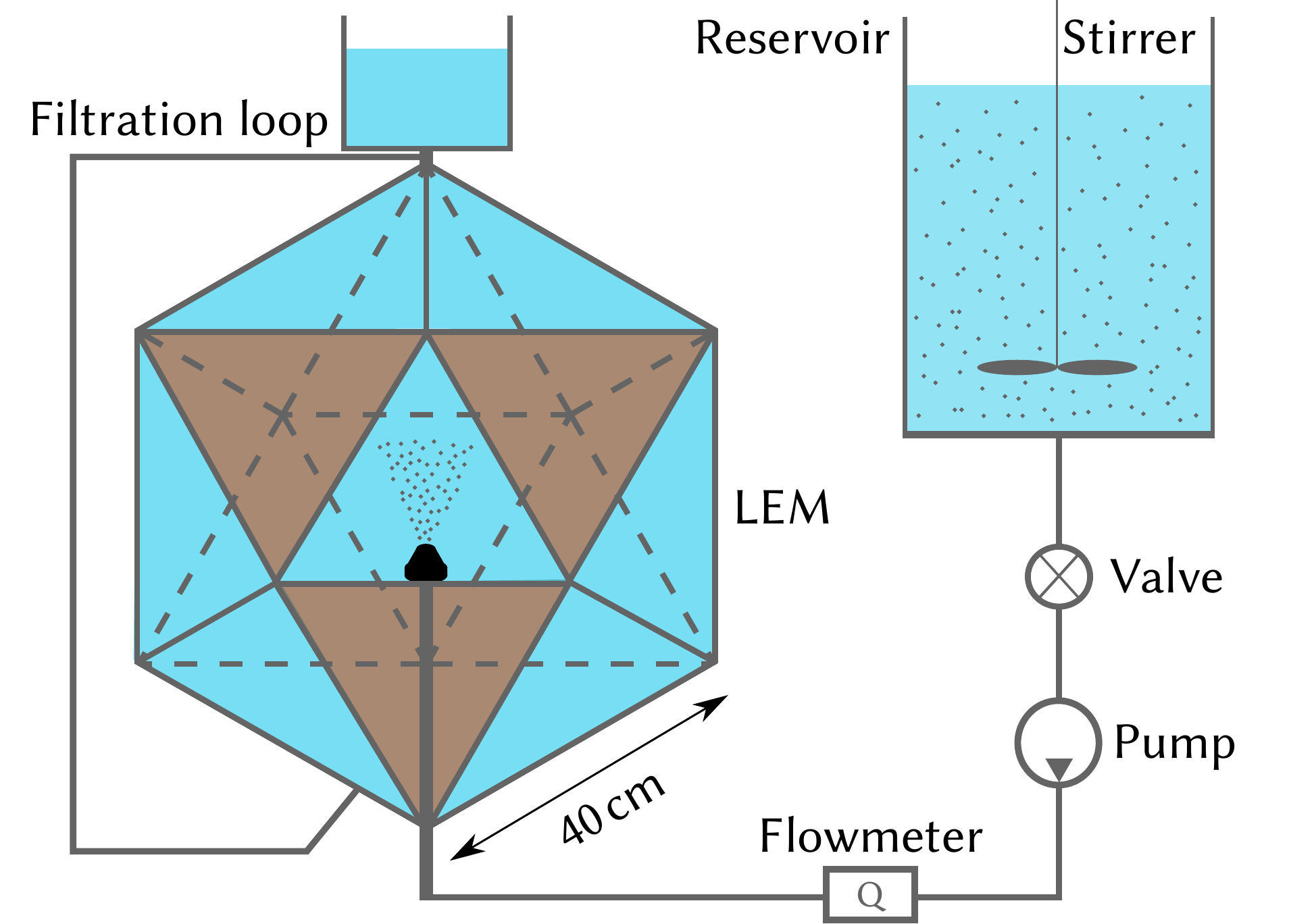}}
\caption{Schematic of the experimental set-up. The three high-speed cameras are oriented orthogonal to the brown faces.\label{fig:setup}}
\end{figure}
A water jet seeded with particles was studied in the Lagrangian Exploration Module (LEM) at the \'Ecole Normale Sup\'erieure de Lyon. The vertical water jet is injected with a pump connected to a reservoir into the LEM, a convex regular icosahedral (20-faced polyhedron) tank full of water. A schematic of the set-up is shown in figure~\ref{fig:setup}. The jet is ejected upwards from a round nozzle with a diameter $D = \SI{4}{mm}$. At the nozzle exit, the flow rate is $Q \simeq \SI{e-4}{m^3/s}$, generating an exit velocity $U_J \simeq \SI{7}{m/s}$, and, in turn, a Reynolds number based on the nozzle diameter $\Rey_D = U_JD/\nu \simeq \SI{2.8e4}{}$ with $\nu$ as the water kinematic viscosity. An overflow valve releases the excess water from the top of the tank at the same rate as injection from the nozzle. Experiments are performed at ambient temperature.

The vertical position of the nozzle is chosen to observe a jet sufficiently far from the walls to discount momentum effects from the LEM onto the jet \citep{hussein1994velocity}, and thus a free jet is observed. The interrogation volume spans $\SI{100}{mm} \leq z \leq \SI{180}{mm}$ ($25 \leq z/D \leq 45$) with the $z$ axis along the jet axis and $z = 0$ the nozzle exit position. In this region, the jet is self-similar (self-similarity holds for $z \gtrsim 15D$) and the centreline velocity is between 1 and $\SI{2}{m/s}$.

The particles, seeding the jet during injection, are neutrally buoyant spherical polystyrene tracers with a density $\rho_p = \SI{1060}{kg/m^3}$ and a diameter $d_p = \SI{250}{\micro\meter}$. The reservoir is seeded with a mass loading of 0.05\% (reasonable seeding to observe a few hundreds of particles per frame) and an external stirrer maintains homogeneity of the particles. The quiescent water inside the LEM is not seeded, therefore tracked particles are, in principle, only those injected into the measurement volume through the nozzle. In practice, it is unavoidable that some few particles eventually end up being resuspended in the surrounding fluid and reentrained within the jet. The probed flow is therefore almost exclusively tagged by nozzle seeded particles with a minor residual contribution of entrained particles. In the following, we will refer to this specific seeding as \textit{nozzle seeding}. Additional measurements with a homogeneous seeding in the whole volume of the LEM (mass loading of 0.1\%) without nozzle seeding are also realised and will be discussed too. The inlet valve is open some seconds before the recording, in such a way that the jet is stationary but minimal particle recirculation occurs, assuring a limited pollution of the surrounding fluid with particles.

Three high-speed cameras (Phantom V12, Vision Research) mounted with $\SI{100}{mm}$ macro lenses (Zeiss Milvus) are used to track the particles. The interrogation volume is illuminated in a back-light configuration with three $\SI{30}{cm}$ square light-emitting diode panels oriented one opposite to each camera. The spatial resolution of each camera is $1280 \times 800$ pixels, creating a measurement volume of around $80 \times 100 \times \SI{130}{mm^3}$. Hence one pixel corresponds to approximately $\SI{0.1}{mm}$. The three cameras are synchronised via TTL triggering at a frequency of $\SI{6}{kHz}$ for 8000 snapshots, resulting in a total record of nearly $\SI{1.3}{s}$ per run. A total of 50 runs are performed to ensure statistical convergence.

\subsection{Particle tracking and post-processing}
Lagrangian particle tracking requires three main steps to compute the tracks: particle detection, stereoscopic reconstruction and tracking. A brief description of the method is presented herein (the particle tracking source codes used for the present study are available on request).
\begin{enumerate}
\item Particle detection enables the measurement of positions of the centres of the particles in the camera images by using an \textit{ad hoc} process based on classical methods of image analysis such as nonuniform illumination correction and centroid detection.
\item Stereoscopic reconstruction aims at finding the particle coordinates in three-dimensional space by combining the two-dimensional views from the three cameras. To achieve this, an accurate calibration is required, allowing the connection of pixel coordinates to real world coordinates. A recent polynomial calibration developed in \citet{machicoane2019simplified} and the matching algorithm by \citet{bourgoin2020using} are used.
\item Tracking of the particles through time transforms the cloud of points into trajectories. This is obtained through classical predictive tracking methods as presented in \citet{ouellette2006quantitative}.
\end{enumerate}

\begin{figure}
\centerline{\includegraphics[width=0.7\textwidth]{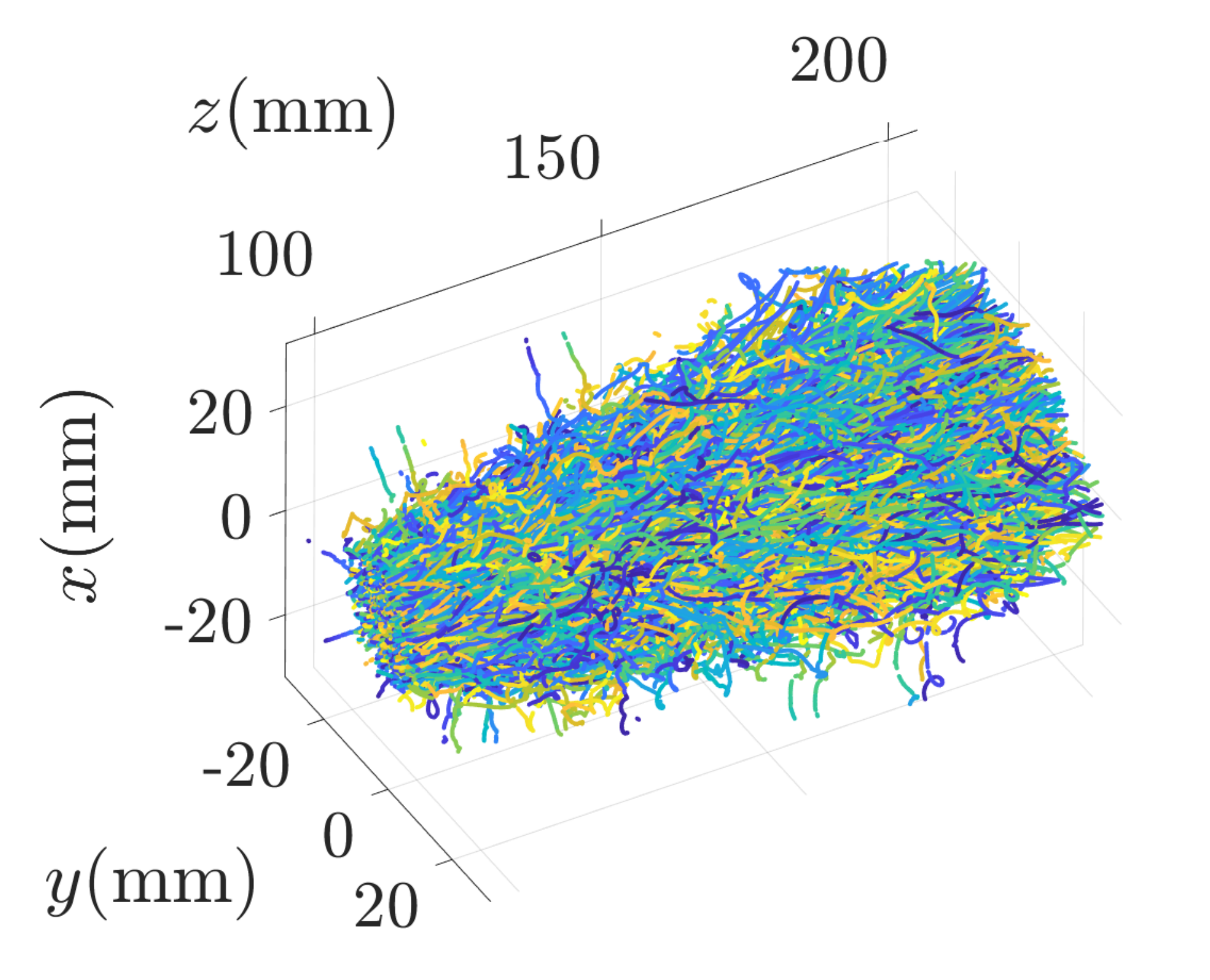}}
\caption{A sample of tracks: 14 182 tracks longer than or equal to 50 frames (one colour per trajectory, one film considered). The majority of the particles come from the nozzle, a few of them come from the tank.\label{fig:jet_tracks}}
\end{figure}
Finally, the trajectories are smoothed by convolution with a Gaussian kernel and the velocities are computed by convolving tracks with a first-order derivative Gaussian kernel \citep{mordant2004experimental}. The coordinate basis is adapted by coinciding the $z$ axis with the jet axis and centring it in $x$ and $y$ directions. Positions and velocities are computed in adapted cylindrical coordinates. A visualisation of tracks is shown in figure~\ref{fig:jet_tracks}. It can be noted that most trajectories come from the nozzle (where they are injected) and very few come from the outside and correspond to particles entrained into the jet (visible in figure~\ref{fig:jet_tracks} as radial trajectories towards the jet). The full data set is comprised of $\SI{3.5e6}{}$ trajectories longer than or equal to 4 frames, which corresponds to $\SI{1.0e8}{}$ particle positions and velocities.

A more complete description of the hydraulic and optical set-ups as well as Lagrangian particle tracking and post-processing methods is given in \citet{viggiano2021lagrangian} which focuses on Lagrangian statistics in the same flow.

\section{Mean velocity field}\label{sec:velocity}
We define the axial velocity $U(z,r,\theta,t)$ with $z$ the axial coordinate, $r$ the radial one, $\theta$ the circumferential one and $t$ the time. We also define the radial velocity $V(z,r,\theta,t)$ and the circumferential velocity $W(z,r,\theta,t)$. The $z$ axis is the jet axis and $z = 0$ is the nozzle exit position (see figure~\ref{fig:jet}\textit{b}). The Eulerian statistics (i.e. time averaged statistics) of these quantities (mean fields, Reynolds stresses, etc.) are well-known through classical Eulerian metrology, such as hot-wire or laser-Doppler anemometry \citep{wygnanski1969measurements, panchapakesan1993turbulence1, hussein1994velocity, pope2000turbulent, lipari2011review}. Time average is denoted $\langle \cdot \rangle$ and time averaged quantities are referred as mean quantities (the studied jet is in stationary state).

In the present study, we focus on the mean axial velocity field $\langle U \rangle (z,r)$ (independent of $\theta$ because of axisymmetry) and the mean radial velocity field $\langle V \rangle (z,r)$ which is smaller than $\langle U \rangle$ by one order of magnitude (the mean circumferential velocity $\langle W \rangle$ is zero due to axisymmetry). We will also investigate in the next section the mean concentration field $\langle \varphi \rangle (z,r)$ of nozzle seeded particles as they spread.

We shall distinguish in the sequel the Eulerian fields of the global jet, $\langle U \rangle$ and $\langle V \rangle$ (which would be measured with a homogeneous seeding), and the fields of the flow solely tagged by nozzle seeded particles, which we denote $\langle U_\varphi \rangle$ and $\langle V_\varphi \rangle$ (other related quantities would also be differentiated from those of the global jet with the subscript $\varphi$).

In practice, these fields are retrieved from the aforementioned Lagrangian experiments, based on nozzle seeded particle trajectories. We consider all particles for all films and all time steps, and bin the measurement volume to compute the mean axial or radial velocity of all particles inside each bin. The resulting fields can be compared to the mean fields from Eulerian measurements. Since the flow is only tagged with nozzle seeded particles, we eventually expect to observe differences between the retrieved velocity field and the Eulerian velocity field of the global jet: $\langle U \rangle \neq \langle U_\varphi \rangle$ and $\langle V \rangle \neq \langle V_\varphi \rangle$.

In the two following subsections, dedicated respectively to the mean axial and radial velocity, we first recall the classical known properties of the mean Eulerian velocity field (compiled in \citet{pope2000turbulent} and \citet{lipari2011review}), then we compare them with those Lagrangian-based measurements.

\subsection{Mean axial velocity}
We first recall known properties of the mean axial velocity in the self-similar region far from the nozzle (approximately for $z \gtrsim 15D$ with $D$ the nozzle diameter). We consider the mean centreline velocity $U_0(z) = \langle U \rangle (z,r=0)$, and its half-width $r_{1/2}(z)$ such that $\langle U \rangle (z,r=r_{1/2}(z)) = U_0(z)/2$. Self-similarity enables characterisation of the mean axial velocity by these three relations:
\begin{equation}
U_0(z) = \dfrac{BU_JD}{z-z_0},
\label{eq:U0}
\end{equation}
with $U_J$ the jet axial velocity at the nozzle, $z_0$ a virtual origin, and $B$ a dimensionless constant (typical values are $z_0 \simeq 4D$ and $B \simeq 5.8$ according to \citet{pope2000turbulent} and \citet{lipari2011review});
\begin{equation}
r_{1/2}(z) = S(z-z_0),
\label{eq:r12}
\end{equation}
with $S$ a dimensionless constant (typical value is $S \simeq 0.094$ according to \citet{pope2000turbulent} and \citet{lipari2011review});
\begin{equation}
f(\eta) = \dfrac{\langle U \rangle (z,r)}{U_0(z)},
\label{eq:f_def}
\end{equation}
which is the radial profile in its self-similar form with the dimensionless self-similar coordinate $\eta = r/(z-z_0)$.

The self-similar mean axial velocity profile $f$ must satisfy some constraints: by definition $f(0) = 1$, while $f'(0) = 0$ because $f$ is even and smooth (the prime notation represents the derivative with respect to the self-similar variable $\eta$). It is also expected to decrease towards 0 as $\eta$ increases (i.e. downstream and/or outwards the jet). However, no exact analytical expression is known for $f$. Because the jet and other free shear flows are \textit{slender} flows, i.e. they do not extend far in the lateral direction and mainly extends in the axial direction, the averaged turbulent boundary-layer equations are the usual theoretical framework for the jet \citep{schlichting2017boundary}. Using these equations as an approximation for the jet dynamics and assuming a constant (uniform) turbulent viscosity \citep{pope2000turbulent, schlichting2017boundary} (it will be further discussed in §~\ref{sec:diff} and appendix~\ref{app:B}), an approximate analytical expression can be calculated for $f$ leading to a squared Lorentzian function:
\begin{equation}
f(\eta) \simeq (1+A\eta^2)^{-2},
\label{eq:f_lorentz2}
\end{equation}
with $A = (\sqrt{2}-1)/S^2$. Experimentally, the squared Lorentzian profile is found to reasonably hold near the jet centreline ($\eta \lesssim 0.15$), but to deviate from the measured profile at larger $\eta$. This indicates that an accurate description of the self-similar mean profile must account for the non-uniformity of the turbulent viscosity, which requires to be experimentally determined. It is empirically found that an improved global fit of $f$ is obtained using a Gaussian function \citep{so1986similarity}:
\begin{equation}
f(\eta) \simeq e^{-A\eta^2},
\label{eq:f_gauss}
\end{equation}
with $A = \log(2)/S^2$.

\begin{figure}
(\textit{a}) \hspace{0.22\textwidth}(\textit{b}) \hspace{0.205\textwidth}(\textit{c}) \hspace{0.205\textwidth}(\textit{d})\\
\includegraphics[width=0.24\textwidth]{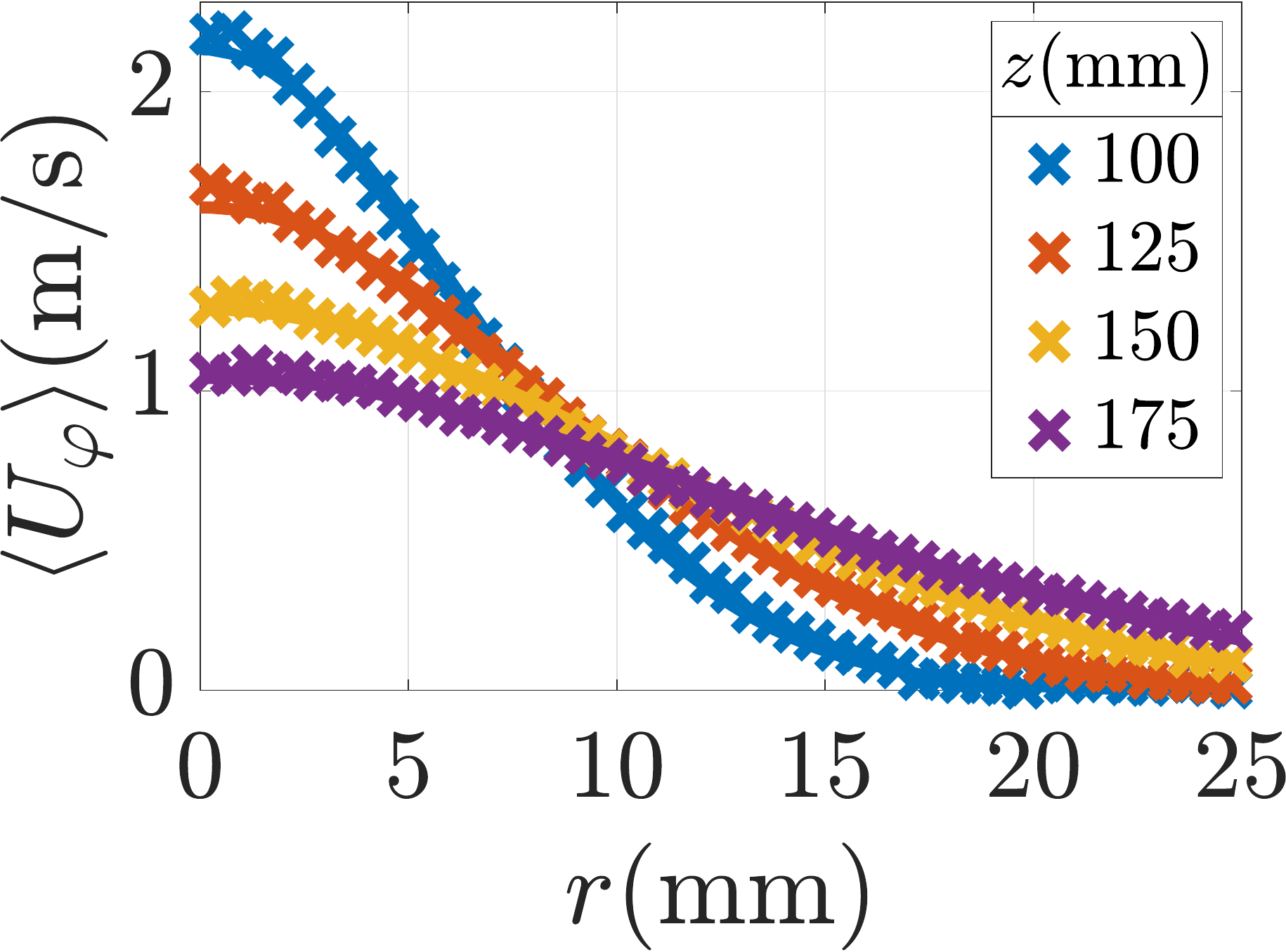}
\includegraphics[width=0.24\textwidth]{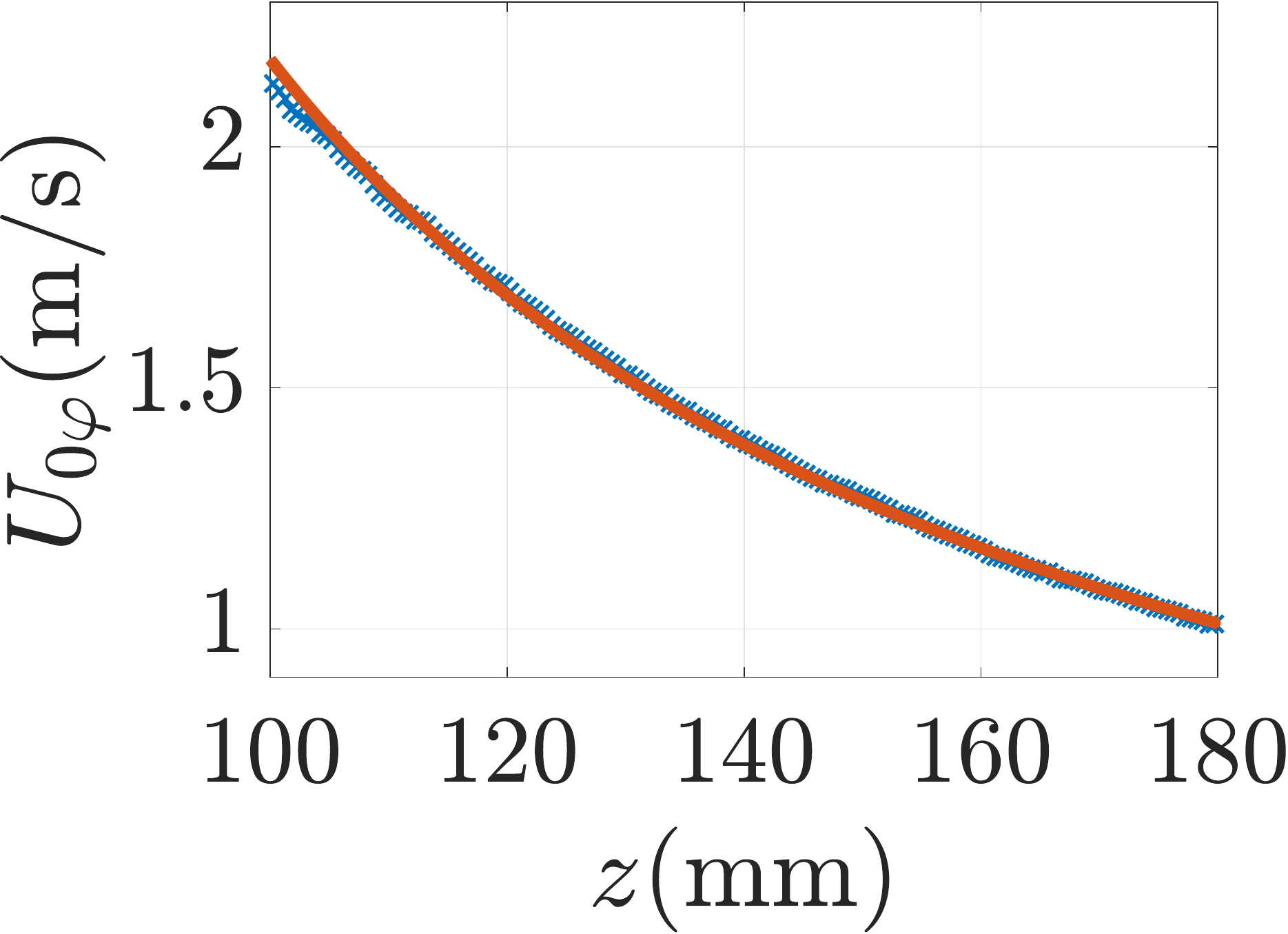}
\includegraphics[width=0.24\textwidth]{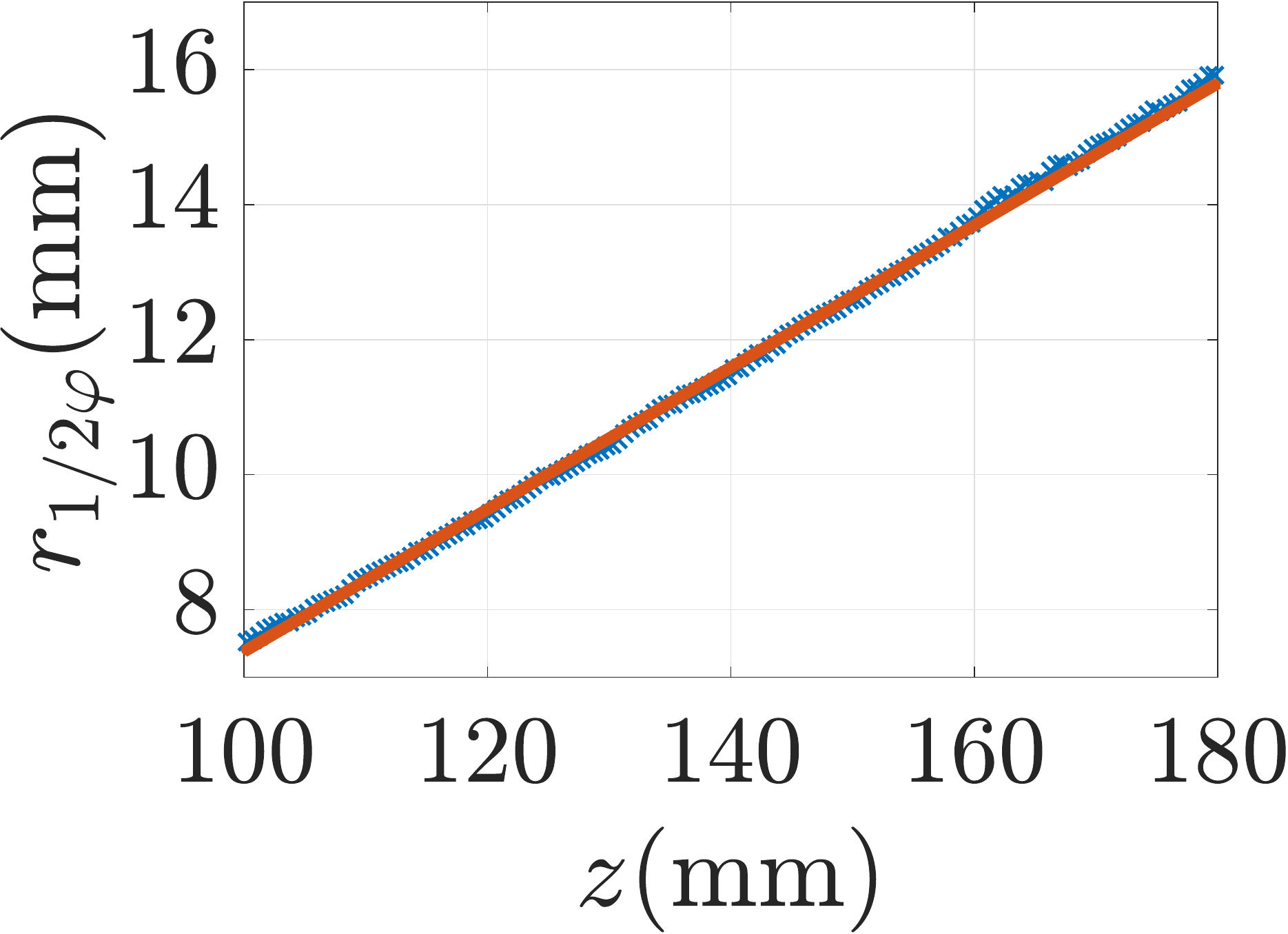}
\includegraphics[width=0.24\textwidth]{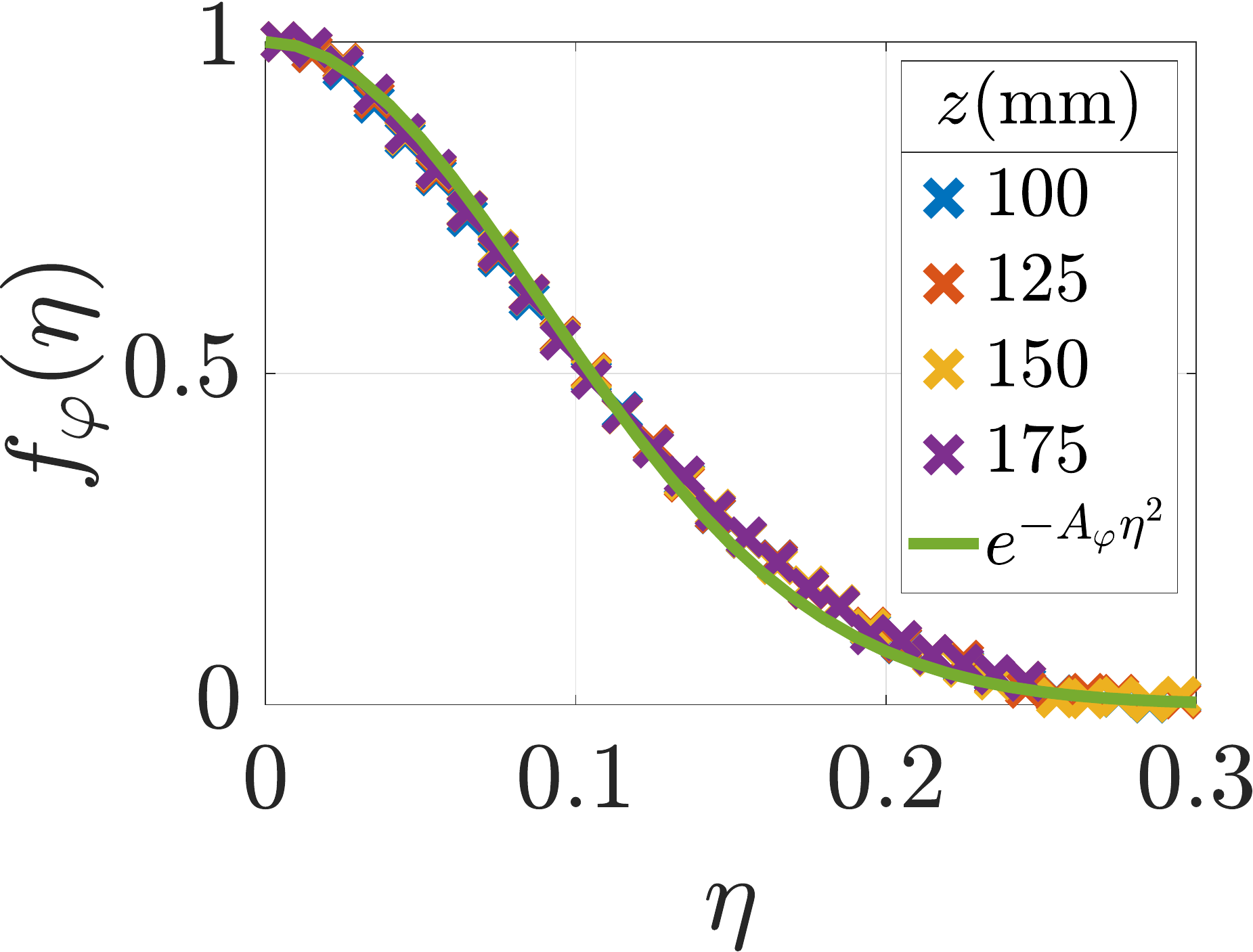}
\caption{Characterisation of the mean axial velocity field $\langle U_\varphi \rangle$ based on trajectories with a nozzle seeding. (\textit{a})~Radial profiles of the mean axial velocity $\langle U_\varphi \rangle$ (crosses: experimental points, solid lines: Gaussian fit). (\textit{b})~Mean centreline velocity $U_{0\varphi}(z)$ (crosses: experimental points, solid line: fit~\eqref{eq:U0}). (\textit{c})~Half-width $r_{1/2\varphi}(z)$ (crosses: experimental points, solid line: fit~\eqref{eq:r12}). (\textit{d})~Self-similar profiles $f_\varphi(\eta)$~\eqref{eq:f_def} (crosses: experimental points, solid line: fit~\eqref{eq:f_gauss}).\label{fig:axial}}
\end{figure}
The estimate of the mean field $\langle U_\varphi \rangle$ (based on experimental trajectories with a nozzle seeding) is performed in cylindrical coordinates $(z,r,\theta)$ and then averaged over $\theta$ (due to axisymmetry) leading to statistics in the two-dimensional space $(r,z)$. In practice, we bin space in $r$ and $z$ every $\SI{0.5}{mm}$ and compute the mean axial velocity of the particles inside each bin. For the self-similar profiles, we bin in $\eta$ by steps of 0.01. Figure~\ref{fig:axial} shows the radial profiles of the mean axial velocity $\langle U_\varphi \rangle (z,r)$ at different downstream positions $z$, the axial evolution of the mean centreline velocity $U_{0\varphi}(z)$ and of the half-width $r_{1/2\varphi}(z)$, and the self-similar profile $f_\varphi(\eta)$ measured in our experiment when probing solely nozzle seeded particles.

When comparing the nozzle seeded particle measurements with the classical Eulerian relations given by~\eqref{eq:U0},~\eqref{eq:r12} and~\eqref{eq:f_def}, we observe an excellent agreement. In particular self-similarity is very well satisfied, with a Gaussian self-similar profile $f_\varphi$ and fitting parameters $B_\varphi = 5.3$ and $S_\varphi = 0.105$ ($A_\varphi = 63$), which are consistent with those classically determined for the global Eulerian jet dynamics \citep{pope2000turbulent, lipari2011review}. The value of $S_\varphi$ is found slightly larger than the values reported in Eulerian measurements which usually span between 0.09 and 0.10 \citep{lipari2011review}, suggesting that the nozzle seeded particle profile is slightly wider than the actual Eulerian profile. Despite this small difference, we will consider in the sequel that $f \simeq f_\varphi$.

This first observation suggests that the axial dynamics of nozzle seeded particles accurately represents the global axial Eulerian dynamics, even if entrained particles are not probed. This reflects that the axial momentum of the jet is primarily associated to nozzle seeded particles. We will see in the next subsection that, on the contrary, entrained particles play a crucial role on the mean radial velocity profile.

\subsection{Mean radial velocity - An incompressibility paradox}
We now perform the same study for the mean radial velocity. As previously done with the mean axial velocity $\langle U \rangle$, we can define a self-similar profile for the mean radial velocity $\langle V \rangle$:
\begin{equation}
g(\eta) = \dfrac{\langle V \rangle (z,r)}{U_0(z)}.
\label{eq:g_def}
\end{equation}
Interestingly in an incompressible jet, $\langle U \rangle$ and $\langle V \rangle$ are linked through the continuity equation
\begin{equation}
\bnabla \bcdot \langle \boldsymbol{U} \rangle = 0,
\label{eq:continuity}
\end{equation}
where $\langle \boldsymbol{U} \rangle = \langle U \rangle \boldsymbol{e_z} + \langle V \rangle \boldsymbol{e_r}$. Combining equation~\eqref{eq:U0} and definitions~\eqref{eq:f_def} and~\eqref{eq:g_def}, the continuity equation~\eqref{eq:continuity} can be rewritten as \citep{pope2000turbulent}
\begin{equation}
\eta (\eta f(\eta))' = (\eta g(\eta))',
\label{eq:continuity_ss}
\end{equation}
which can be integrated to obtain the following general relation between the self-similar mean radial and axial profiles for the global Eulerian dynamics of an incompressible free round jet:
\begin{equation}
g(\eta) = \eta f(\eta) - \dfrac{1}{\eta} \int_0^\eta x f(x) \:\mathrm{d}x.
\label{eq:g_gen}
\end{equation}
Knowing that $f(0) = 1$ and $f'(0) = 0$, we deduce that $g(0) = 0$, $g'(0) = 1/2$ and $g''(0) = 0$. Using the empirical Gaussian approximation~\eqref{eq:f_gauss} for $f$, equation~\eqref{eq:g_gen} gives the following approximated expression for $g$:
\begin{equation}
g(\eta) \simeq \eta e^{-A\eta^2} - \dfrac{1-e^{-A\eta^2}}{2A\eta}.
\label{eq:g_gauss}
\end{equation}

\begin{figure}
\centerline{\includegraphics[width=0.5\textwidth]{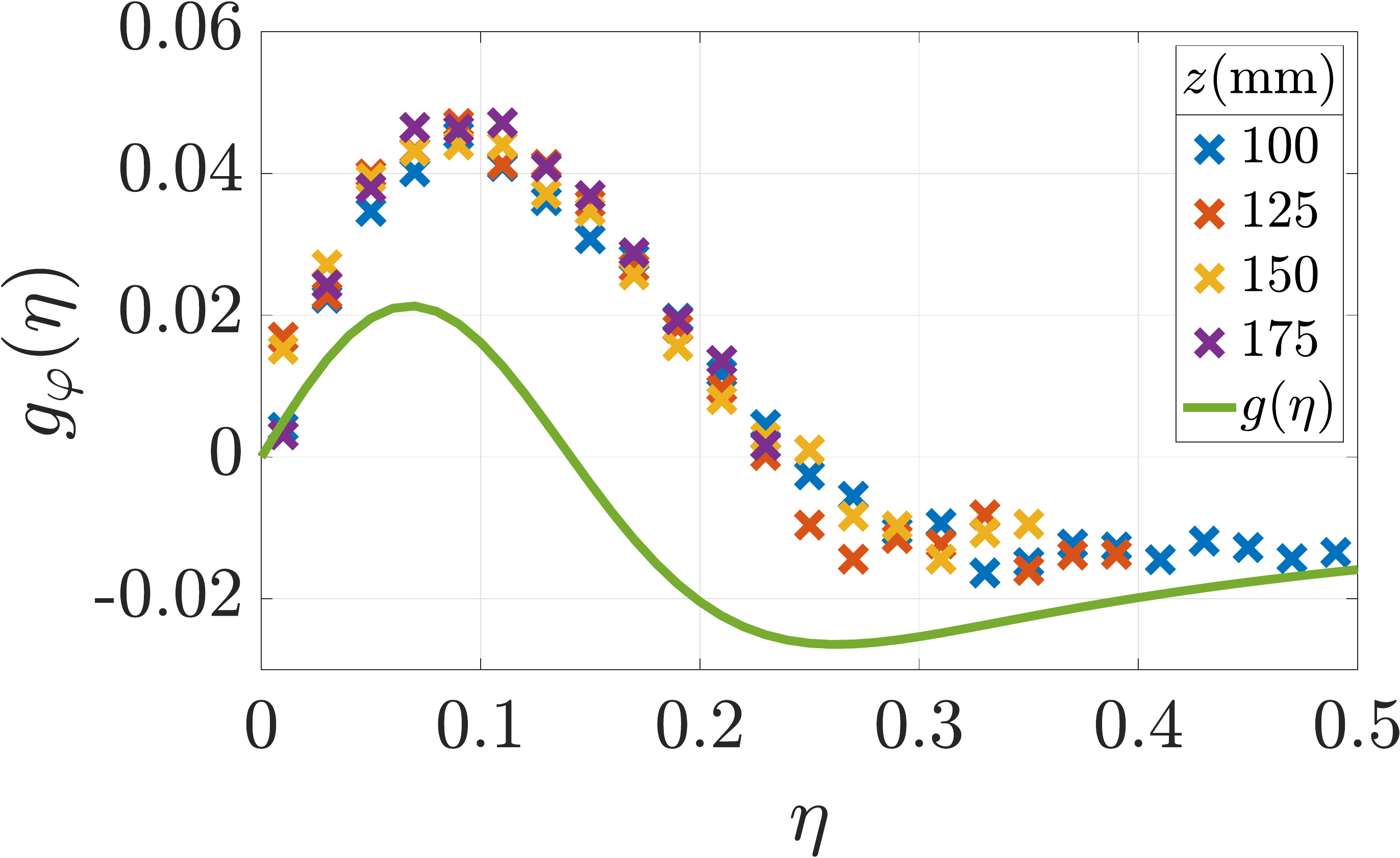}}
\caption{Self-similar profiles $g_\varphi(\eta)$~\eqref{eq:g_def} for a nozzle seeding (crosses: experimental points, solid line: fit~\eqref{eq:g_gauss} with $A_\varphi = 63$ previously found for $f_\varphi(\eta)$).\label{fig:radial}}
\end{figure}
Figure~\ref{fig:radial} presents the experimental mean radial velocity profile $g_\varphi(\eta)$ for the nozzle seeding case (obtained as for the axial velocity, binning $z$ in steps of $\SI{0.5}{mm}$ and $\eta$ in steps of 0.02), which is compared to the self-similar profile $g(\eta)$~\eqref{eq:g_gauss} expected for $\langle V \rangle$ from the previous incompressibility considerations for the global Eulerian profile. It can be observed that, though the measured profiles of $g_\varphi$ do hold self-similarity, they strongly deviate from the expected self-similar incompressible profile for the global jet $g$. More specifically, three points can be highlighted: (i)~the amplitude of the measured maximums of $g_\varphi$ is twice that of the expected incompressible profile $g$, (ii)~the measured profiles cross zero at a much higher value of $\eta$, and (iii)~the slope at the origin ($\eta = 0$) of the measured self-similar profile is 1 instead of 1/2.

Overall, contrary to the mean axial velocity profile which is essentially indistinguishable between the nozzle seeding case and the global Eulerian field ($f_\varphi \simeq f$), the mean radial velocity profile is strongly affected by the nozzle seeding up to the core of the jet ($g_\varphi \neq g$). Since the radial and axial velocity profiles are classically linked by simple incompressibility considerations (as just discussed), and considering that the jet under investigation does operate in incompressible conditions, this discrepancy may appear at first sight as a paradox.

\begin{figure}
\centerline{\includegraphics[width=0.5\textwidth]{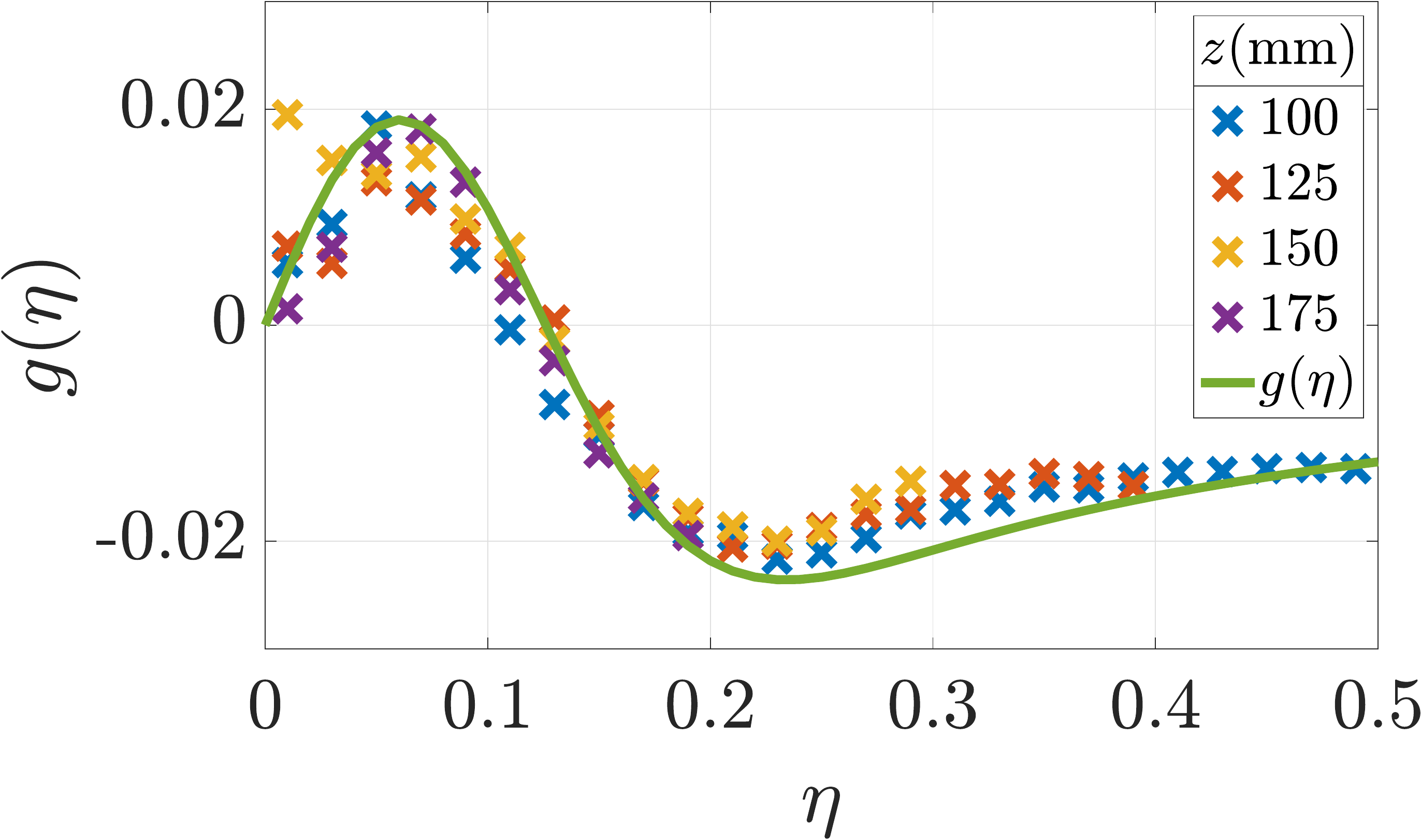}}
\caption{Self-similar profiles $g(\eta)$ for a homogeneous seeding in the whole volume of the LEM without nozzle seeding (crosses: experimental points, solid line: fit~\eqref{eq:g_gauss} with $A = 79$).\label{fig:radial_homo}}
\end{figure}
In order to rule out any possible experimental error at the origin of the major difference observed between the measured profile with a nozzle seeding $g_\varphi$ and the expected global incompressible profile $g$, we performed experiments with an actual homogeneous seeding in the whole volume of the tank. The measured radial profile $g(\eta)$, shown in figure~\ref{fig:radial_homo}, accurately matches the expected incompressible profile~\eqref{eq:g_gauss}. Some discrepancy can be observed for $\eta \gtrsim 0.2$, which can be attributed to the fact that $f$ is less well fitted by a Gaussian function as it decreases towards zero. Moreover, with this homogeneous seeding, we find $S = 0.094$ which is a usual value for $S$ \citep{lipari2011review}.

The radial profile $g_\varphi$ therefore truly appears to not comply with global incompressibility constraint. Section~\ref{sec:velocity_phi} presents a simple theoretical explanation (based on mass conservation) of this apparent paradox, which quantitatively describes the experimental observations through an effective compressibility of the velocity field associated to nozzle seeded particles. The physical origin of this effective compressibility relies on the role played by entrained particles (not accounted for when only nozzle seeded particles are tracked).

Before presenting these theoretical developments, we briefly discuss the qualitative role played by entrained particles on the observed discrepancies between $g_\varphi$ and $g$. The strong influence of the seeding (nozzle or homogeneous) on the radial profiles can indeed be qualitatively interpreted from the specific role played by entrained particles (not tagged with a nozzle seeding). Particles entrained from outside to inside the jet require a negative radial velocity to reach the core of the jet and therefore contribute negatively to the global radial velocity profile $g$, while particles injected through the nozzle and subsequently spreading outwards are mainly associated to a positive radial contribution to $g$. Therefore, when a homogeneous seeding is considered, the combination of these two contributions (spreading and entrainment) eventually leads to a global radial profile $g$ (see figure~\ref{fig:radial_homo}), where spreading dominates in the centre ($g(\eta) > 0$ for $\eta < 0.13$) and entrainment dominates on the sides ($g(\eta) <0$ for $\eta > 0.13$). When only nozzle seeded particles are tagged, the role of entrained particles is not accounted for. As a consequence, an overall hindering of the negative radial contribution associated to those particles is expected, leading to a higher and mostly positive profile for $g_\varphi$, as experimentally measured (see figure~\ref{fig:radial}).

We present in the next section a simple theoretical and quantitative description of the combined role of spreading and entrainment based on the mean tracer concentration.

\section{Effective compressibility and entrainment}\label{sec:velocity_phi}
We qualitatively explained the differences between $g$ and $g_\varphi$ by the absence of the contribution due to entrained particles in $g_\varphi$. We also pointed that, considering that $f \simeq f_\varphi$ and that the expected $g$ as expressed in equation~\eqref{eq:g_gauss} comes directly from incompressibility considerations, the discrepancy between $g_\varphi$ and $g$ implies that the measured mean velocity field $\langle \boldsymbol{U_\varphi} \rangle$ associated to nozzle seeded particles behaves as compressible, i.e. $\bnabla \bcdot \langle \boldsymbol{U_\varphi} \rangle \neq 0$. This is at first sight in contradiction with the experimental conditions as the free jet under investigation is actually incompressible. The apparent compressibility of the flow tagged solely by nozzle seeded particles is actually a simple consequence of the inhomogeneous seeding (as presented in figure~\ref{fig:radial_homo}, with a homogeneous seeding in the whole experimental volume, the retrieved velocity profiles do comply with incompressibility). In this section, we rationalise this effective compressibility, giving an explicit relation between $g$ and $g_\varphi$ which emphasises the contribution of entrained particles.

\subsection{Nozzle seeding model}
To account for effective compressibility and compute $g_\varphi$, we propose to generalise the classical approach relating mean radial and axial velocity profiles through incompressibility, in order to account for the inhomogeneity of the concentration field (itself due to the inhomogeneous seeding). 

We denote $\varphi(z,r,\theta,t)$ the instantaneous concentration field of nozzle seeded tracers. As we did for the mean axial and radial velocities, we consider the mean concentration field $\langle \varphi \rangle (z,r)$. The continuity equation for the mean concentration field $\langle \varphi \rangle$ and the mean velocity field $\langle \boldsymbol{U_\varphi}\rangle$ imposes that
\begin{equation}
\bnabla \bcdot (\langle \varphi \rangle \langle \boldsymbol{U_\varphi} \rangle) = 0.
\label{eq:continuity_phi}
\end{equation}
Note that, because by definition $\boldsymbol{U_\varphi}$ is exactly the advection velocity of the nozzle seeded tracers (not including any eventually unknown random velocity perturbation, $\boldsymbol{U_\varphi}$ is not a Eulerian field), the continuity equation as written above for the mean (concentration and velocity) fields is exact, as there is no additional diffusion term associated to the transport of the tracers by the unperturbed advection velocity $\boldsymbol{U_\varphi}$. Note also that for a homogeneous seeding (i.e. $\langle \varphi \rangle$ independent of all spatial coordinates), equation~\eqref{eq:continuity_phi} naturally reduces to the classical incompressible relation $\bnabla \bcdot \langle \boldsymbol{U_\varphi} \rangle = 0$, which however does not hold when $\langle \varphi \rangle$ is inhomogeneous, as for the case of nozzle seeded tracers investigated here.

\begin{figure}
\hspace{0.15\textwidth}(\textit{a}) \hspace{0.3\textwidth}(\textit{b})\\
\centerline{\includegraphics[width=0.3\textwidth]{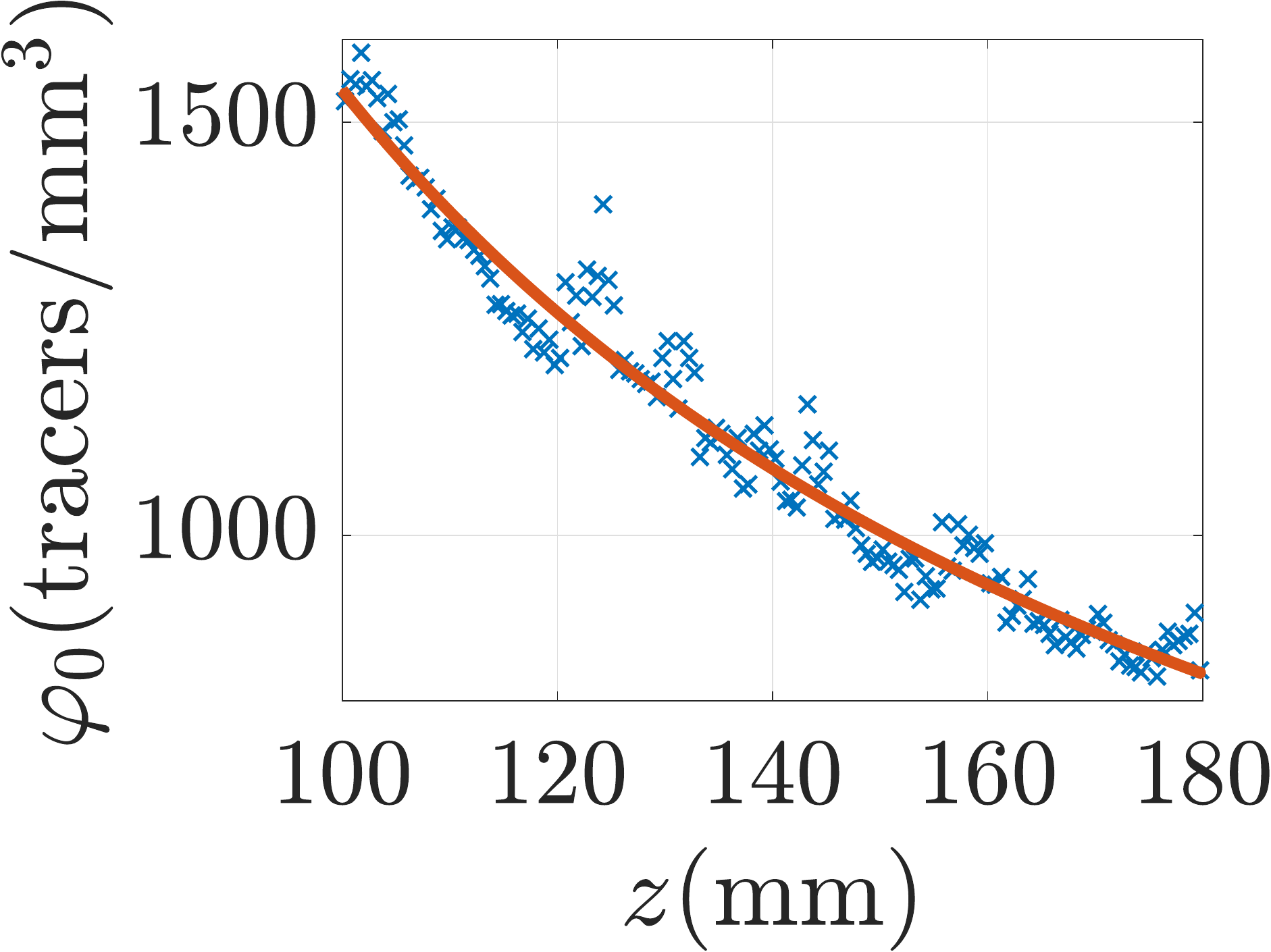}
\includegraphics[width=0.3\textwidth]{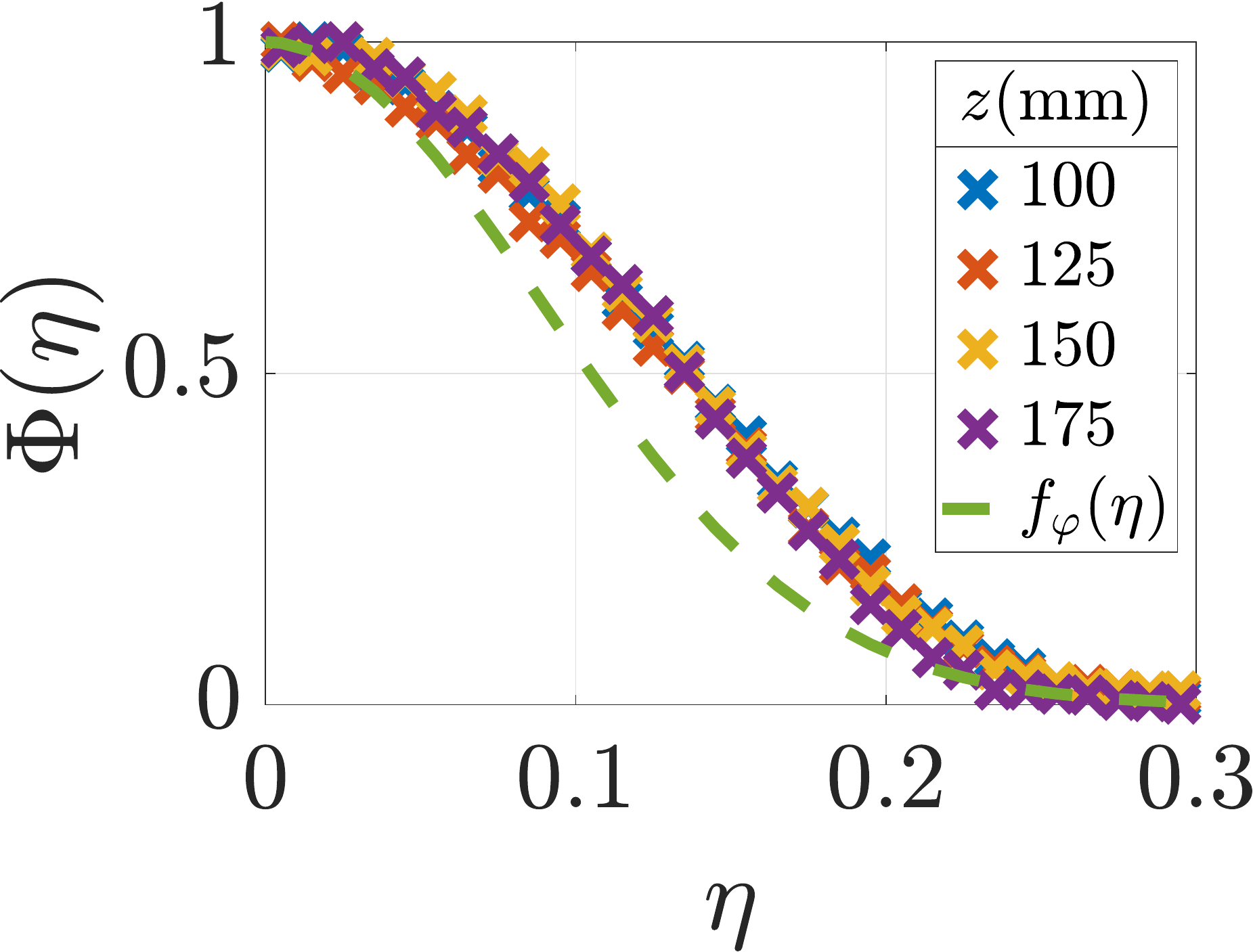}}
\caption{Characterisation of the mean concentration field $\langle \varphi \rangle$. (\textit{a})~Centreline concentration $\varphi_0(z)$ (crosses: experimental points, solid line: fit in $1/(z-z_0)$). $\varphi_0$ is the sum of the concentrations from all films at all time steps, which explains the high values of $\varphi_0$, but only the relative evolution along $z$ is relevant. (\textit{b})~Self-similar profiles $\Phi(\eta)$~\eqref{eq:phi} (crosses: experimental points, dashed line: $f_\varphi(\eta)$ previously measured). The profiles of $\Phi(\eta)$ are wider than those of $f_\varphi(\eta)$.}\label{fig:phi}
\end{figure}
To solve equation~\eqref{eq:continuity_phi}, we first characterise the mean concentration field $\langle \varphi \rangle (z,r)$. Figure~\ref{fig:phi} shows the main properties of $\langle \varphi \rangle$: the mean centreline concentration $\varphi_0(z)$ evolves as $1/(z-z_0)$ and we can define a self-similar profile
\begin{equation}
\Phi(\eta) = \dfrac{\langle \varphi \rangle (z,r)}{\varphi_0(z)},
\label{eq:phi}
\end{equation}
with $\varphi_0(z) \propto 1/(z-z_0)$. The fact that $\langle \varphi \rangle$ evolves as $\langle U \rangle$ can be justified by the behaviour of a conserved passive scalar in a jet. Actually, it is known that, because the boundary-layer equations for the mean axial velocity $\langle U \rangle$ and a scalar field $\langle \varphi \rangle$ are similar, a conserved passive scalar scales with $z$ in the same way as the mean axial velocity does, and the self-similar profile is similar, usually wider (see \citet{pope2000turbulent}). For the present concentration field, the profiles of $\Phi$ are wider than those of $f$, this difference of width and also the shape of $\Phi$ will be discussed in the next section.

From equation~\eqref{eq:continuity_phi} and definition~\eqref{eq:phi}, we infer that self-similar profiles of mean concentration, radial and axial velocity of nozzle seeded particles must satisfy the following relation:
\begin{equation}
\Phi(\eta) [(\eta g_\varphi(\eta))' - \eta (\eta f_\varphi(\eta))'] + \eta [g_\varphi(\eta) \Phi'(\eta) - f_\varphi(\eta)(\eta \Phi(\eta))'] = 0,
\label{eq:continuity_phi_ss}
\end{equation}
which simplifies to
\begin{equation}
g_\varphi(\eta) = \eta f_\varphi(\eta).
\label{eq:g_phi_gen}
\end{equation}
The details of this calculation are given in appendix~\ref{app:A}. It can be noticed that this result does not depend on the exact shape of $\Phi$: only the dependence of $\varphi_0(z)$ in $1/(z-z_0)$ and the self-similarity of $\Phi(\eta)$ are required.

Interestingly, the solution for the effectively compressible fields in the case of the nozzle seeding turns out to be somehow simpler than the global incompressible case, as it does not carry the additional term 
\begin{equation}
\zeta(\eta) = - \dfrac{1}{\eta} \int_0^\eta x f(x) \:\mathrm{d}x.
\label{eq:zeta}
\end{equation}
Going back to equation~\eqref{eq:g_gen} and considering $f = f_\varphi$, we can see that the global mean radial velocity profile (accounting for both nozzle seeded and entrained particles) can be written as the sum of the profile of the nozzle seeded particles alone and this $\zeta$ term:
\begin{equation}
g = g_\varphi + \zeta.
\label{eq:g_gphi_zeta}
\end{equation}
The $\zeta$ contribution can therefore be interpreted as the effect of entrained particles on the global mean radial velocity profile of the jet. Its negative sign naturally reflects the inward flux of particles due to entrainment. Therefore, we will refer to $\zeta$ as the \textit{entrainment term}.

\subsection{Experimental validation}
\begin{figure}
\centerline{\includegraphics[width=0.5\textwidth]{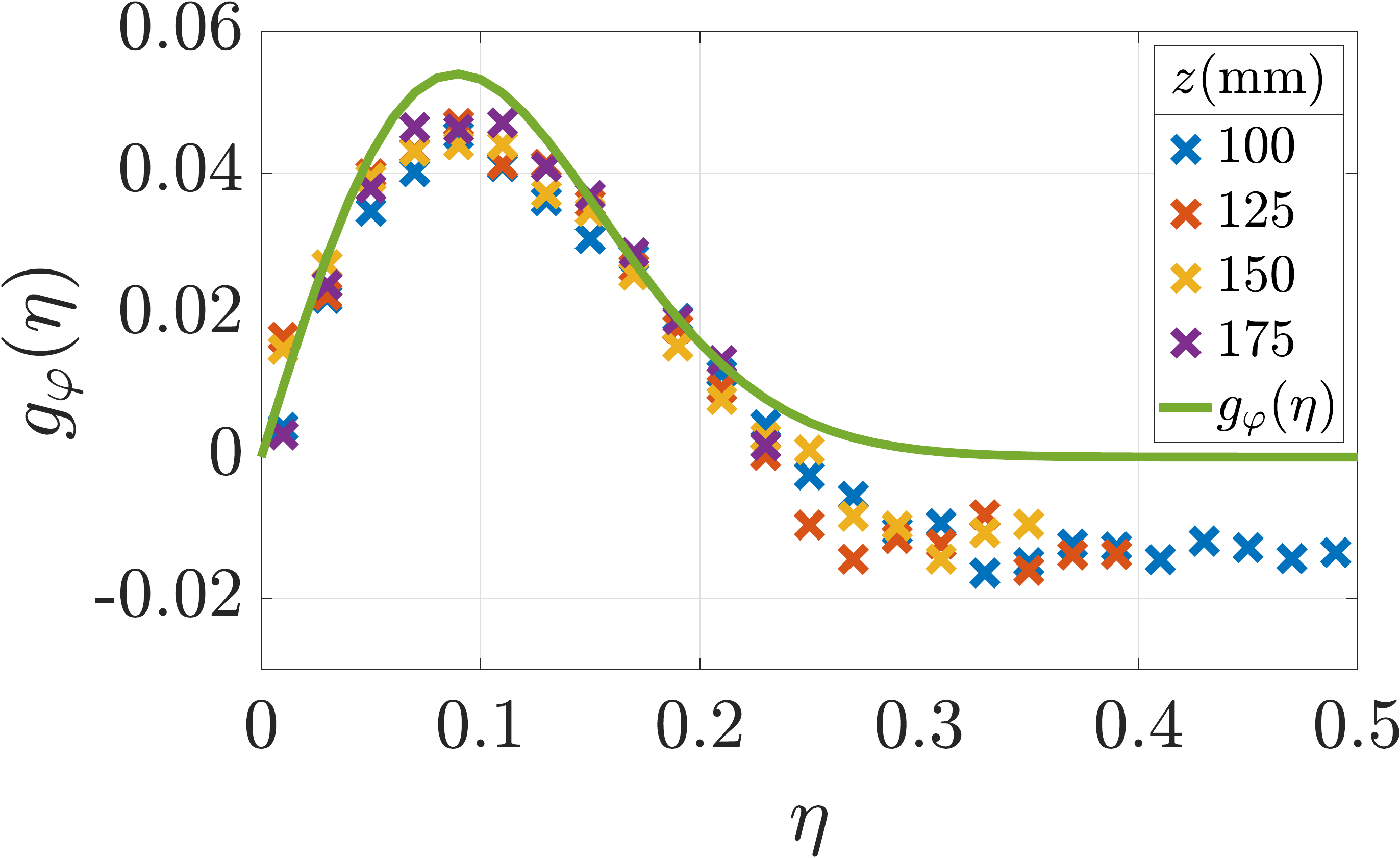}}
\caption{Self-similar profiles $g_\varphi(\eta)$ for a nozzle seeding (crosses: experimental points, solid line: fit~\eqref{eq:g_phi_gauss} with $A_\varphi = 63$ previously found for $f_\varphi(\eta)$). This is the same figure as figure~\ref{fig:radial} but with the new fit~\eqref{eq:g_phi_gauss}.\label{fig:radial_phi}}
\end{figure}
A first interesting property of equation~\eqref{eq:g_phi_gen} is that as $f_\varphi(0) = 1$ by definition, then $g'_\varphi(0) = 1$. This is agreement with the experimental slope of 1 observed in figure~\ref{fig:radial} for $g_\varphi(\eta)$ at $\eta = 0$. Considering a Gaussian function for $f_\varphi$, which was found in previous section to reasonably matches the experimental measurements, we have the expression
\begin{equation}
g_\varphi(\eta) \simeq \eta e^{-A\eta^2}.
\label{eq:g_phi_gauss}
\end{equation}
Figure~\ref{fig:radial_phi} compares this expression to the experimental profiles for $g_\varphi$, showing a much better agreement than the usual expression tested in figure~\ref{fig:radial} for the global profile $g$, with not only the expected slope at the origin, but also a reasonable overall shape, at least up to $\eta \lesssim 0.2$. The main noticeable difference concerns the negative part of the experimental $g_\varphi$ for the largest values of $\eta$, while the prediction given by equation~\eqref{eq:g_phi_gauss} remains positive. This negative part reflects the presence of an inward radial velocity in the outer regions of the jet. This is very likely to be attributed to the presence of few remaining particles in the ambient fluid (not injected at the nozzle) been entrained into the core of the jet. As a consequence, if some entrained particles are indeed tagged, it is expected that the radial profile measured is not exactly $g_\varphi$ but also carries some contribution due to the negative entrainment term $\zeta$. These few entrained particles with negative radial velocity may also explain the slight overestimation of the maximum of the radial velocity profile prediction compared to the experimental data. Despite this bias, experimental data globally supports the validity of relation~\eqref{eq:g_phi_gauss} and hence of~\eqref{eq:g_phi_gen}.

\begin{figure}
\hspace{0.1\textwidth}(\textit{a}) \hspace{0.42\textwidth}(\textit{b})\\
\centerline{\includegraphics[width=0.8\textwidth]{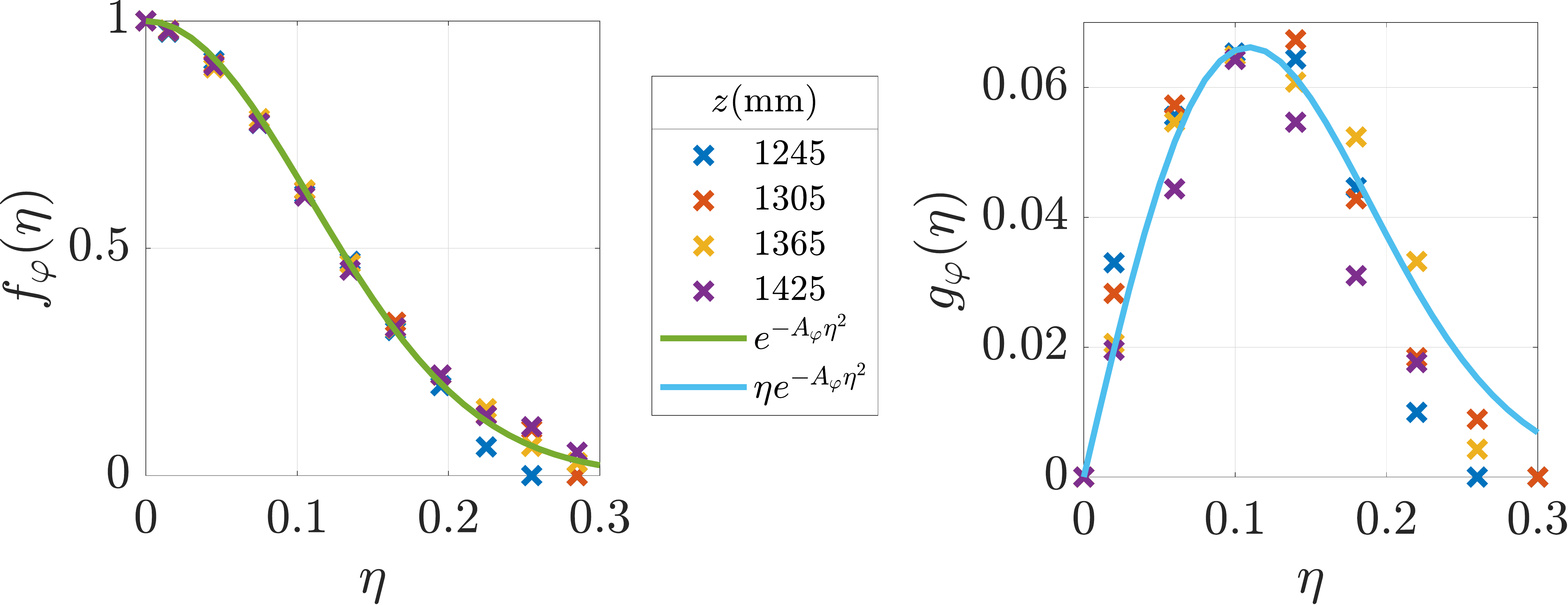}}
\caption{Characterisation of the mean velocity field for an air jet seeded trough the nozzle with neutrally buoyant soap bubbles. Self-similar profiles for mean (\textit{a})~axial and (\textit{b})~radial velocities (crosses: experimental points, solid lines: fits~\eqref{eq:f_gauss} and~\eqref{eq:g_phi_gauss} with $A_\varphi = 42$).\label{fig:LEGI_results}}
\end{figure}
The validity of these relations is also tested on a separate data set from an independent experiment, using similar methods at the Universit\'e Grenoble Alpes with a self-similar round free air jet seeded with neutrally buoyant millimetric soap bubbles inflated with helium ($D = \SI{2.25}{cm}$, $U_J \simeq \SI{25}{m/s}$, $\Rey_D \simeq \SI{3.7e4}{}$, $d_p = \SI{2.5}{mm}$). The advantage of this set-up is that the jet blows in a very large room, and that helium filled soap bubbles have a finite life time, so that experiments can be run with the warranty that no spurious particles remain in the ambient fluid surrounding the jet. Mean axial and radial velocity profiles for this experiment are represented in figure~\ref{fig:LEGI_results}. The statistical convergence of this new data set is not as accurate as for the water experiment and the accessible measurement volume does not allow to explore values of $\eta$ above 0.3. However, it can still be seen that no negative values of $g_\varphi$ are measured and that the maximum of the experimental profile matches very well the predicted in that case where entrained particles have been totally avoided. The slight difference in the profiles between the air and water experiments (for instance the maximum of $g_\varphi$ in air is a bit larger than in water) are related to a slightly different value of the fitting parameter $A_\varphi$ of the Gaussian fit for the mean axial velocity profile $f_\varphi$, which could be linked to different geometries of the setup or to the total absence of entrained particles in the air jet.

\section{Link with turbulent diffusion}\label{sec:diff}
\begin{figure}
\centerline{\includegraphics[width=0.7\textwidth]{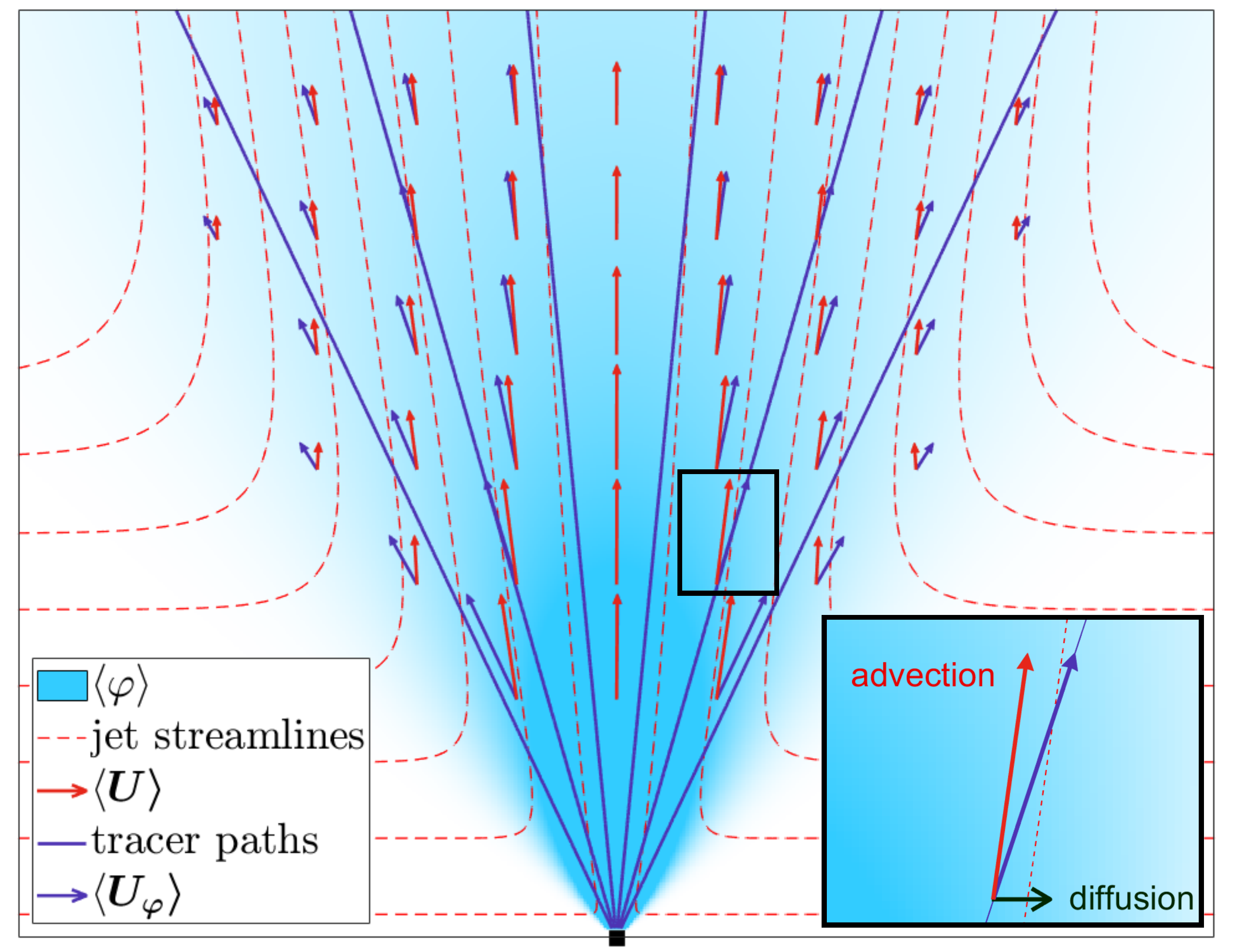}}
\caption{Schematic of the nozzle seeding case with $\langle \boldsymbol{U_\varphi} \rangle = \langle \boldsymbol{U \rangle} + \langle \boldsymbol{U_d} \rangle$. The colour scale represents the tracer concentration $\langle \varphi \rangle$. A first set of streamlines (dashed lines) is used to represent the mean trajectories of the fluid parcels with the associated velocity field $\langle \boldsymbol{U} \rangle$. A second set of streamlines (solid lines) represents the mean trajectories of the tracers coming from the nozzle with the associated velocity field $\langle \boldsymbol{U_\varphi} \rangle$. Excepted on the axis of the jet, the streamlines of the tracers differ from the jet streamlines due to the inhomogeneous nozzle seeding. It results that $\langle \boldsymbol{U} \rangle$ and $\langle \boldsymbol{U_\varphi} \rangle$ have the same axial component but different radial components. This difference can be related to a transverse diffusive flow $\langle \boldsymbol{U_d} \rangle$, as represented in the inset.\label{fig:diff_model}}
\end{figure}
Classical mean field approaches to describe the spreading of substances or particles in turbulent flows usually rely on advection-diffusion modelling for the mean concentration profile. In such approaches the mean transport of the spreading particles is considered to result from two contributions: the advection by the mean velocity $\langle \boldsymbol{U} \rangle$ of the surrounding turbulence and a diffusive velocity $\langle \boldsymbol{U_{d}} \rangle$ modelling the mean field effect of unresolved small scale fluctuations. In such a framework, the mean velocity field of the transported substance $\langle \boldsymbol{U_\varphi} \rangle$ can therefore be written as $\langle \boldsymbol{U_\varphi} \rangle = \langle \boldsymbol{U \rangle} + \langle \boldsymbol{U_d} \rangle$. This is schematically represented in figure~\ref{fig:diff_model}. In the previous section, we showed that the difference between the global mean velocity field $\langle \boldsymbol{U} \rangle$ and the actual mean velocity field $\langle \boldsymbol{U_\varphi} \rangle$ of nozzle seeded particles is related to the entrainment mechanism through the entrainment term $\zeta$ via mass conservation: $\zeta$ ensures the incompressibility of the global field (including both, the entrained and nozzle seeded particles), while the nozzle seeded particle velocity $\langle \boldsymbol{U_\varphi} \rangle$ is effectively compressible. The equivalence of these two approaches (advection/diffusion and global flow/entrainment) to describe the spreading of nozzle seeded particles suggests that the diffusive contribution in the former shall therefore be itself related to the entrainment contribution in the latter.

The aim here is to link these two fields, $\langle \boldsymbol{U} \rangle$ and $\langle \boldsymbol{U_\varphi} \rangle$, through the mean concentration field of particles $\langle \varphi \rangle$ as previously presented in figure~\ref{fig:phi}, with an advection-diffusion model, in order to explicitly connect turbulent diffusion and entrainment.

\subsection{Advection-diffusion equation with turbulent diffusivity $K_T$}
We consider that the tracers are, on one hand, advected by the mean flow, and on the other hand, spread by turbulence. Modelling this turbulent process as diffusive, we write
\begin{equation}
\bnabla \bcdot (\langle \varphi \rangle \langle \boldsymbol{U} \rangle - K_T \bnabla \langle \varphi \rangle) = 0,
\label{eq:continuity_K}
\end{equation}
with $K_T$ the turbulent diffusivity. Equation~\eqref{eq:continuity_K} is the same as equation~\eqref{eq:continuity_phi} with the relation between $\langle \boldsymbol{U} \rangle$ and $\langle \boldsymbol{U_\varphi} \rangle$
\begin{equation}
\langle \boldsymbol{U_\varphi} \rangle = \langle \boldsymbol{U} \rangle - K_T \dfrac{\bnabla \langle \varphi \rangle}{\langle \varphi \rangle},
\label{eq:U_phi_K}
\end{equation}
where $\langle \boldsymbol{U_d} \rangle = -K_T \dfrac{\bnabla \langle \varphi \rangle}{\langle \varphi \rangle}$ represents the aforementioned diffusive contribution.

With previous definitions for the self-similar mean axial and radial velocity fields and mean concentration profile, and considering the decay law for the centreline velocity from~\eqref{eq:U0} ($U_0(z) = BU_JD/(z-z_0)$), equation~\eqref{eq:U_phi_K} leads to these two expressions for the self-similar mean axial and radial velocity profiles of the spreading particles:
\begin{align}
& f_\varphi(\eta) = f(\eta) + \dfrac{K_T(\eta)}{BU_JD} \left[1+\eta\dfrac{\Phi'(\eta)}{\Phi(\eta)}\right],
\label{eq:f_phi_K}\\
& g_\varphi(\eta) = g(\eta) - \dfrac{K_T(\eta)}{BU_JD} \dfrac{\Phi'(\eta)}{\Phi(\eta)},
\label{eq:g_phi_K}
\end{align}
where the first term in the right-hand side of both expressions accounts for advection and the second for diffusion. At this stage these two equations~\eqref{eq:f_phi_K} and~\eqref{eq:g_phi_K} are nothing but mathematical expressions reflecting the \textit{a priori} advection/diffusion decomposition of the particle velocity in~\eqref{eq:U_phi_K}. To be physically relevant, they have to be consistent with the experimental observations and the results of the mass conservation presented in previous sections for $f$, $g$, $f_\varphi$ and $g_\varphi$. 

First, our experiments show that $f \simeq f_\varphi$. To be consistent with~\eqref{eq:f_phi_K}, this requires the second term of this relation to be negligible compared to $f$. Experimental measurements of the turbulent diffusivity $K_T$ and of the self-similar mean concentration field $\Phi$ (presented in the following) confirm the validity of this approximation (this term has the same order of magnitude than $g$, thus it is more than one order of magnitude smaller than $f$).

Second, to be consistent with~\eqref{eq:g_gphi_zeta}, equation~\eqref{eq:g_phi_K} implies that
\begin{equation}
K_T(\eta) = - BU_JD \dfrac{\Phi(\eta)}{\Phi'(\eta)} \dfrac{1}{\eta} \int_0^\eta x f(x) \:\mathrm{d}x.
\label{eq:K_gen}
\end{equation}
Thus the turbulent diffusivity $K_T(\eta)$ is a self-similar quantity dependent on space and expression~\eqref{eq:K_gen} gives a practical relation to estimate it from the knowledge of simple mean field quantities (namely mean concentration and mean axial velocity profiles) which are easily measurable. This contrasts both with classical simplistic approaches assuming a constant turbulent diffusivity and with the usual fundamental definition of turbulent diffusivity, based on the cross-correlation between velocity and concentration fluctuations \citep{pope2000turbulent}.

$K_T(\eta)$ as given by relation~\eqref{eq:K_gen} is a dimensional quantity (with units $\SI{}{m^2/s}$). Similarly to all other self-similar quantities characterising the jet, and as it is done for turbulent viscosity, a dimensionless turbulent diffusivity $\widehat{K}_T$ can be defined:
\begin{equation}
\widehat{K}_T(\eta) = K_T(\eta) / (U_0(z)r_{1/2}(z)) = - \dfrac{1}{S} \dfrac{\Phi(\eta)}{\Phi'(\eta)} \dfrac{1}{\eta} \int_0^\eta x f(x) \:\mathrm{d}x,
\label{eq:K_hat_gen}
\end{equation}
which can ultimately be rewritten as
\begin{equation}
\widehat{K}_T(\eta) = \frac{\zeta(\eta)}{S\chi(\eta)},
\label{eq:K_hat_balance}
\end{equation}
where $\zeta(\eta) = -\dfrac{1}{\eta} \displaystyle \int_0^\eta x f(x) \:\mathrm{d}x$ has already been defined in~\eqref{eq:zeta} and shown to be associated to entrainment, $\chi(\eta)= \Phi'(\eta)/\Phi(\eta)$ characterises the persistent inhomogeneity of the seeding and can be interpreted as a compressibility factor associated to the flow of nozzle seeded particles, and $S = \tan(\delta) \simeq \delta$ with $\delta$ the semi opening angle of the jet cone based on $r_{1/2}$.

Overall, relation~\eqref{eq:K_hat_balance} synthesises the connection between the \textit{a priori} advection/diffusion mathematical decomposition of particle velocity and the physical considerations of mass conservation developed in previous sections by connecting the turbulent diffusivity $K_T$ to (i)~entrainment (via $\zeta$), (ii)~apparent compressibility of the dispersing phase (via $\chi$), and (iii)~global spreading of the jet (via $S$). Note that a conceptually similar connection between effective diffusivity and effective compressibility has been proposed in the context of mixing in linear flows \citep{raynal2018advection}.

\subsection{Turbulent diffusivity and turbulent viscosity}
The turbulent diffusivity $K_T$ and the turbulent viscosity $\nu_T$ are both effective transport coefficients defined in the framework of a mean field description (transport of mass for the first and of momentum for the second). They model the average contribution of small scale turbulence via cross-correlation terms of fluctuating quantities ($\langle u\varphi' \rangle$ for $K_T$ and $\langle uv \rangle$ for $\nu_T$, with fluctuating quantities $u = U - \langle U \rangle$, $v = V - \langle V \rangle$ and $\varphi' = \varphi - \langle \varphi \rangle$ \citep{pope2000turbulent}). This formal analogy between $K_T$ and $\nu_T$, together with the importance of $\nu_T$ for practical numerical modelling strategies (such as RANS approaches) and the simplicity of the relations established in the previous subsection allowing the estimation of $K_T$ from simple measurements of mean field quantities, motivate us to further extend previous considerations (connecting turbulent diffusivity to entrainment and mass conservation) in order to revisit formal links between turbulent diffusivity and turbulent viscosity.

The relation between $K_T$ and $\nu_T$ is commonly written in terms of the turbulent Prandtl number, $\sigma_T = \nu_T/K_T$, which compares the efficiency of momentum and mass transport. Several studies have investigated the turbulent Prandtl number by studying for instance the turbulent transport of conserved passive scalars such as temperature \citep{corrsin1950further, chevray1978intermittency, chua1990turbulent, ezzamel2015dynamical} or concentration of chemical species \citep{papanicolaou1988investigations, dowling1990similarity, panchapakesan1993turbulence2, lemoine1996simultaneous, chang2002turbulent}, leading to values of $\sigma_T$ of the order of unity (experimental values around 0.7 are usually reported). However, there is no consensus about how $\sigma_T$ exactly depends on space and none of these studies explicitly address the question of a possible formal connection with simple mean field quantities.

\subsubsection{Uniform $\sigma_T$}
In the case where $\sigma_T$ is assumed to be uniform (independent of space), it can be shown from the turbulent boundary-layer equations (see \citet{schlichting2017boundary}) that
\begin{equation}
\Phi(\eta) = f(\eta)^{\sigma_T} \quad \textrm{or equivalently} \quad \sigma_T = \frac{\log\Phi}{\log f}.
\label{eq:sigma_uniform}
\end{equation}
This relation combined with the expression of $K_T$~\eqref{eq:K_hat_gen} leads to the following expression for the turbulent viscosity:
\begin{equation}
\widehat{\nu}_T(\eta) = - \dfrac{1}{S} \dfrac{f(\eta)}{f'(\eta)} \dfrac{1}{\eta} \int_0^\eta x f(x) \:\mathrm{d}x.
\label{eq:nu_hat_gen}
\end{equation}
As for $K_T$, $\nu_T$ can be inferred by simply measuring the profile $f$ of mean axial velocity and is analytically connected to the entrainment term $\zeta$.

If we consider for instance a squared Lorentzian approximation~\eqref{eq:f_lorentz2} for $f$, expression~\eqref{eq:nu_hat_gen} simplifies to a constant value:
\begin{equation}
\widehat{\nu}_T^\mathrm{Lorentz} = \dfrac{S}{8(\sqrt{2}-1)}.
\label{eq:nu_hat_lorentz2} 
\end{equation}
This is expected, as the squared Lorentzian profile for $f$ is known to be the exact solution of the turbulent boundary-layer equations for a constant turbulent viscosity \citep{pope2000turbulent} (what is experimentally reasonable for $\eta \lesssim 0.15$). Besides, the relation found in equation~\eqref{eq:nu_hat_lorentz2} between $\widehat{\nu}_T$ and $S$ coincides with the classical result when solving the boundary-layer equations for a constant turbulent viscosity.

Expression~\eqref{eq:nu_hat_gen} is however more general and remains valid beyond the constant turbulent viscosity approximation (it still requires the turbulent Prandtl number to be constant though). In particular, if the Gaussian approximation~\eqref{eq:f_gauss} is considered for $f(\eta)$ (which is empirically known to better match the experimental self-similar profiles), the following space-dependent profile is retrieved for the turbulent viscosity:
\begin{equation}
\widehat{\nu}_T^\mathrm{Gauss}(\eta) = \dfrac{S}{4\log(2)} \dfrac{1-e^{-A\eta^2}}{A\eta^2}.
\label{eq:nu_hat_gauss}
\end{equation}
This result is not new, and has been previously derived by \citet{so1986similarity} who propose a generalisation of the solution of the turbulent boundary-layer equations for a non-uniform turbulent viscosity. By considering different experimental functions used to fit $f$, they argue that the Gaussian function is the best one to fit experimental profiles of $f$ and they analytically determine the expression for $\widehat{\nu}_T$ for a Gaussian function, which is exactly the same as equation~\eqref{eq:nu_hat_gauss}.

At this point, we have therefore shown that formula~\eqref{eq:sigma_uniform} (valid in the case of a uniform turbulent Prandtl number $\sigma_T$) allows us to extend the connection established in the previous subsection, between turbulent diffusivity and entrainment, to turbulent viscosity with relation~\eqref{eq:nu_hat_gen}. Besides, this quite general relation is found in agreement with previous derivations, based on boundary-layer equations, for squared Lorentzian and Gaussian mean axial velocity profile. Next subsection generalises formula~\eqref{eq:sigma_uniform} to the case of non-uniform $\sigma_T$.

\subsubsection{Generalisation to non-uniform $\sigma_T$}
In appendix~\ref{app:B}, we show that the general equations~\eqref{eq:K_hat_gen} and~\eqref{eq:nu_hat_gen} for $\widehat{K}_T(\eta)$ and $\widehat{\nu}_T(\eta)$, respectively, relating the self-similar profiles of turbulent diffusivity and turbulent viscosity to the self-similar profiles of mean concentration $\Phi$, mean axial velocity $f$ and entrainment term $\zeta$ are actually the general solutions of the boundary-layer equations.

Furthermore, we also conclude that these two relations remain valid even if the turbulent Prandtl number $\sigma_T(\eta)$ is not constant, and we show that
\begin{equation}
\sigma_T(\eta) = \dfrac{\Phi'(\eta)}{\Phi(\eta)} \dfrac{f(\eta)}{f'(\eta)},
\label{eq:sigma_nonuniform}
\end{equation}
generalisation of formula~\eqref{eq:sigma_uniform}.

Altogether, beyond the conceptual interest of relating effective transport coefficients in the jet to the entrainment process, relations~\eqref{eq:K_hat_gen},~\eqref{eq:nu_hat_gen} and~\eqref{eq:sigma_nonuniform} are of great practical interest as they allow determination of the spatial profiles of turbulent diffusivity, turbulent viscosity and turbulent Prandtl number from the simple measurements of the mean axial velocity profile and the mean concentration profile without requiring the measurement of second-order correlations.

In the next subsection, we apply these relations to experimental measurements.

\subsection{Experimental determination of $K_T$, $\nu_T$ and $\sigma_T$}
\begin{figure}
\centerline{\includegraphics[width=0.5\textwidth]{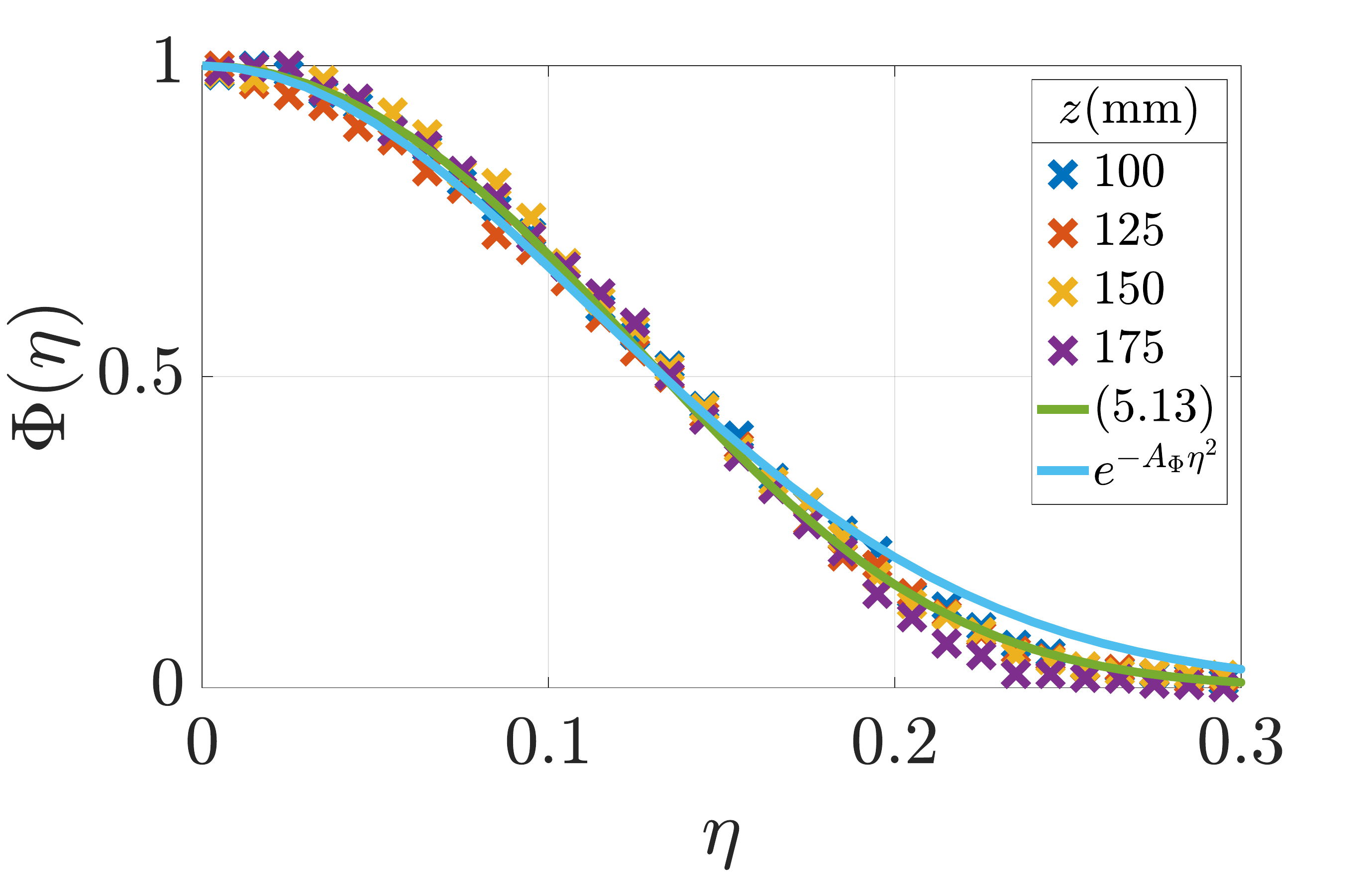}}
\caption{Self-similar profiles $\Phi(\eta)$ (crosses: experimental points, solid lines: fit~\eqref{eq:phi_erf} and Gaussian fit with $A_\Phi = 39$). This is the same figure as figure~\ref{fig:phi}(\textit{b}) but with the new fit~\eqref{eq:phi_erf}.\label{fig:phi_erf}}
\end{figure}
According to equations~\eqref{eq:K_hat_gen},~\eqref{eq:nu_hat_gen} and~\eqref{eq:sigma_nonuniform}, $\widehat{K}_T$, $\widehat{\nu}_T$ and $\sigma_T$ can be experimentally determined from the sole knowledge of the profiles of $f$ and $\Phi$ (besides, only $f$ is required to determine $\widehat{\nu}_T$). As these relations include the derivatives of $f$ and $\Phi$, instead of using the raw experimental profiles, it is useful to consider functional fits of these, which can be more easily manipulated.
\begin{itemize}
\item As already discussed, and as it can be observed in figure~\ref{fig:axial}(\textit{d}), $f$ is reasonably fitted by a Gaussian function. However, for a better accuracy, we use the fitting function $f(\eta) = e^{-a\eta^2}(1+c_2\eta^2+c_4\eta^4)$ introduced by \citet{hussein1994velocity} to fit their experimental measurement of $f(\eta)$ (they also use similar functions to fit the Reynolds stresses). This Gaussian function corrected by a polynomial, although less practical, is closer to the experimental points and leads to a more accurate estimate, in particular, of the derivative $f'(\eta)$ which appears in the formula~\eqref{eq:nu_hat_gen} for the turbulent viscosity. The polynomial correction has a minor impact on the estimate of the integral entrainment term $\zeta$.
\item As it can be observed in figure~\ref{fig:phi_erf}, the concentration profile $\Phi(\eta)$ is broader than a Gaussian function for small values of $\eta$ (typically $\eta<0.1$) and steeper than a Gaussian function for large values of $\eta$. We empirically find that a better function to fit $\Phi(\eta)$ is
\begin{equation}
\Phi(\eta) = \dfrac{\erf((\eta+a)/b)-\erf((\eta-a)/b)}{2\erf(a/b)},
\label{eq:phi_erf}
\end{equation}
(green line in figure~\ref{fig:phi_erf}, to be compared to the Gaussian fit in light blue), where $\erf(x) = 2/\sqrt{\pi} \int_0^x e^{-t^2} \:\mathrm{d}t$ is the error function and $a$ and $b$ the parameters of the fit (here $a = 0.126$ and $b = 0.102$).
\end{itemize}

\subsubsection{Determination of $K_T$}
\begin{figure}
\centerline{\includegraphics[width=0.5\textwidth]{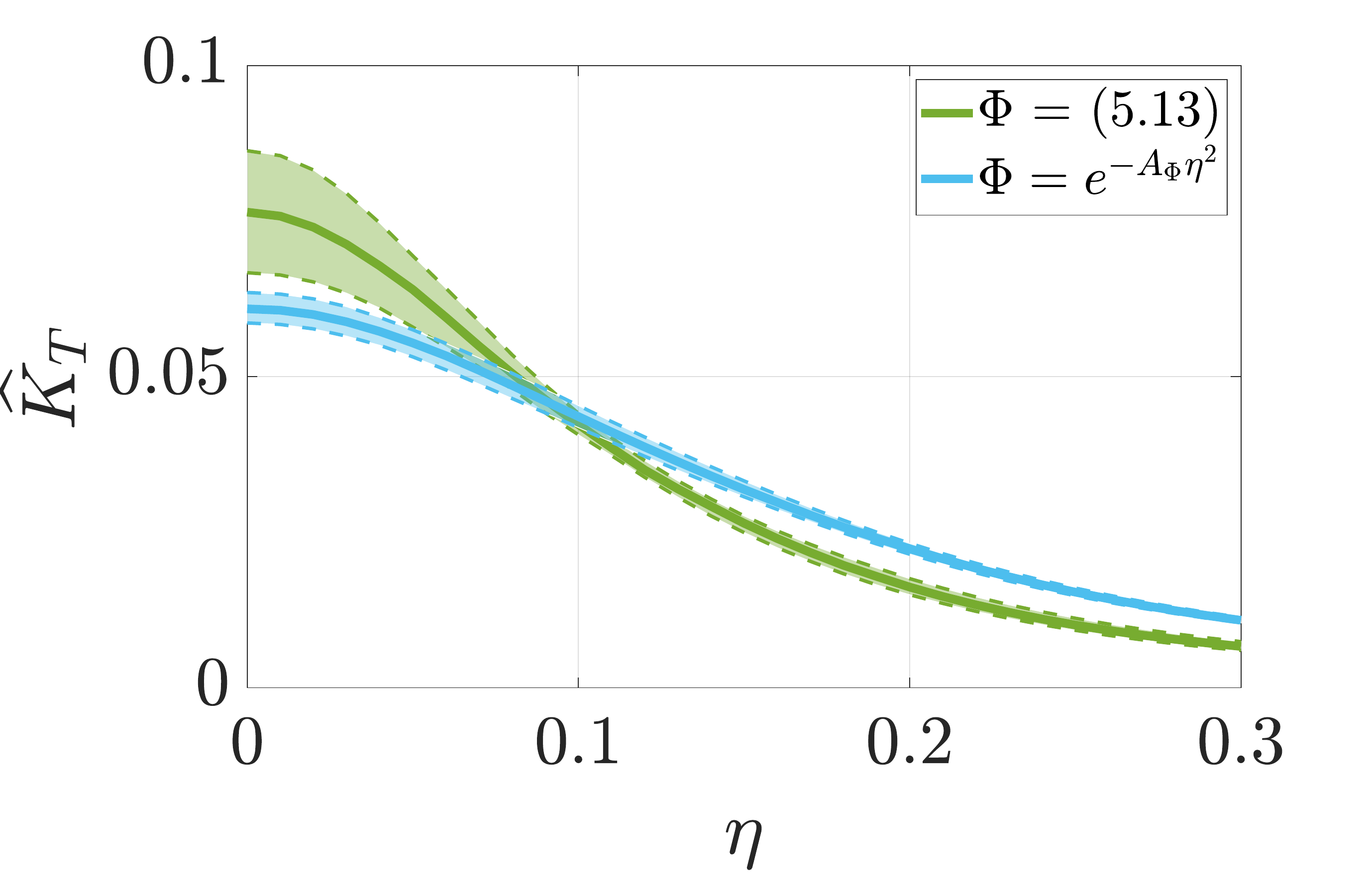}}
\caption{Self-similar profile $\widehat{K}_T(\eta)$ based on two fits of $\Phi$ (solid lines: median values, coloured zones limited by dashed lines: 70\% of the measured values).\label{fig:K}}
\end{figure}
Based on these fits for $f$ and $\Phi$, we compute the experimental profiles of $\widehat{K}_T(\eta)$ from~\eqref{eq:K_hat_gen}, which are shown in figure~\ref{fig:K}. Profiles are obtained for measurements at different streamwise distances from the nozzle between $z = \SI{100}{mm}$ and $z = \SI{180}{mm}$. The solid line is the median value for all $z$ positions along the axis, and the coloured zone between the two dashed lines comprises 70\% of the measured values. The profile of $\widehat{K}_T$ based on a Gaussian fit of $\Phi$ is also represented for comparison, showing that small differences between the two fitting functions for $\Phi$ lead to large differences for estimate of $\widehat{K}_T$. A good determination of the profile of $\widehat{K}_T(\eta)$ therefore requires an accurate measurement of $\Phi(\eta)$. Figure~\ref{fig:K} indicates that the sensitivity to the fit is particularly crucial near the centreline. This can be rationalised from~\eqref{eq:K_hat_gen}, from which it can be shown that $\widehat{K}_T(0) = -1/(2S\Phi''(0))$: the centreline value of $\widehat{K}_T(\eta)$ is related to the curvature at the origin of $\Phi(\eta)$. This explains the underestimate of $\widehat{K}_T(0)$ from the Gaussian fit, which is narrower than the error function fit~\eqref{eq:phi_erf}. It also explains the higher variability of the estimate of $\widehat{K}_T$ from the error function fit near the centreline when data from all axial distances $z$ are considered. Indeed, figure~\ref{fig:phi_erf} shows that although very good, self-similarity is not perfect within the accessible range of distance from nozzle ($z/D \leq 45$). In particular, a mild variation of the curvature at the origin of $\Phi(\eta)$ measured at different downstream distances $z$ can be seen. This sensitivity to small deviations from self-similarity becomes however marginal away from the centreline. Overall, and in spite imperfect self-similarity effects near the centreline (what can be expected to be improved in future studies exploring distances beyond $z/D > 45$), figure~\ref{fig:K} shows that a reasonable profile of $\widehat{K}_T$ can indeed be retrieved from~\eqref{eq:K_hat_gen} only requiring the determination of mean concentration and axial velocity profiles. Few of such measurements of radial inhomogeneity of turbulent diffusivity are available in the literature, mainly due to the complexity of requiring simultaneous measurements of velocity and scalar fluctuations, as classical estimates are based on velocity-scalar cross-correlations. The profile of $\widehat{K}_T$ in figure~\ref{fig:K} is in good agreement with such previous measurements in round free jets \citep{chua1990turbulent, lemoine1996simultaneous, chang2002turbulent}.

\subsubsection{Determination of $\nu_T$}
\begin{figure}
(\textit{a}) \hspace{0.42\textwidth}(\textit{b})\\
\includegraphics[width=0.45\textwidth]{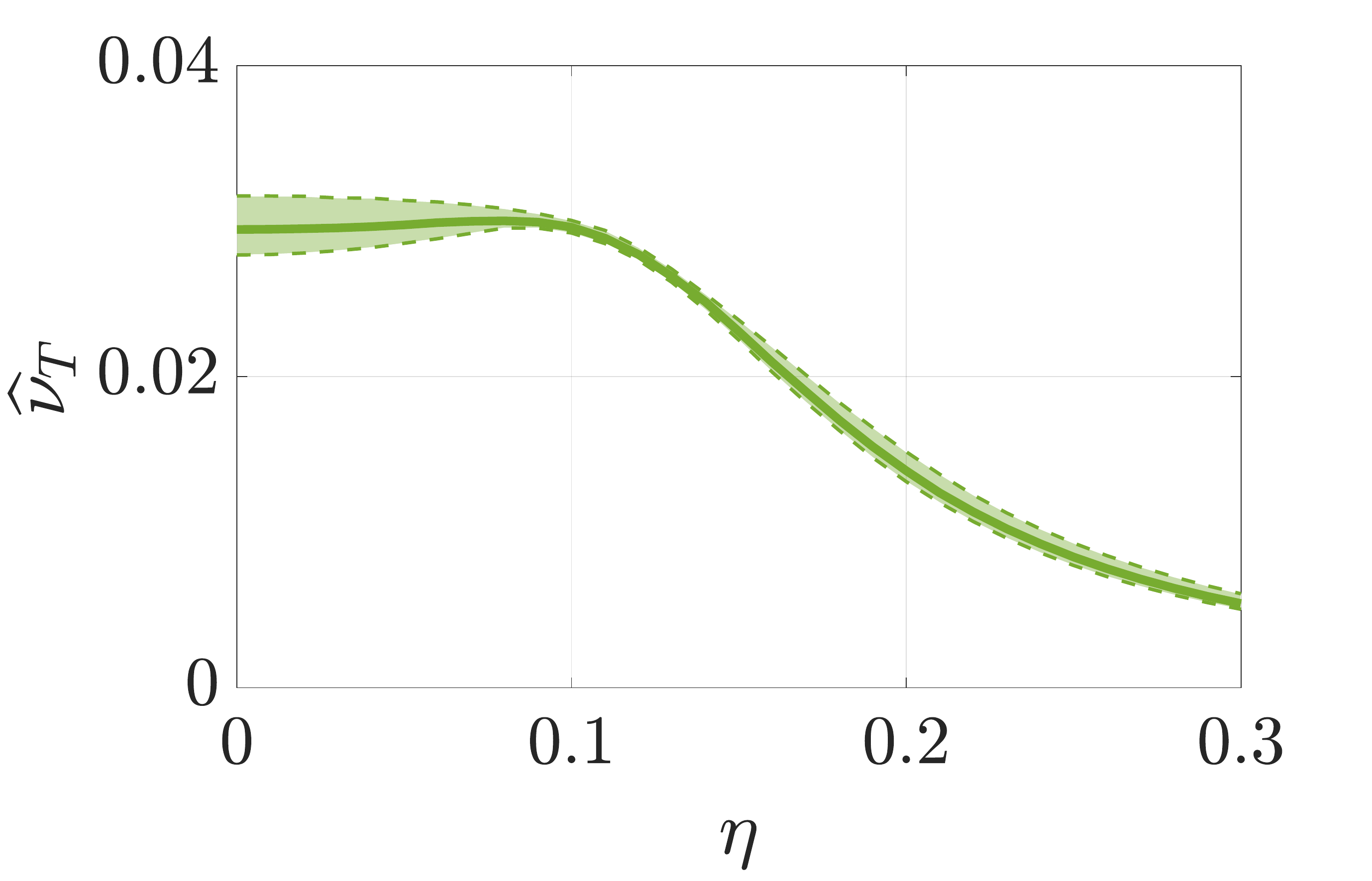}
\includegraphics[width=0.45\textwidth]{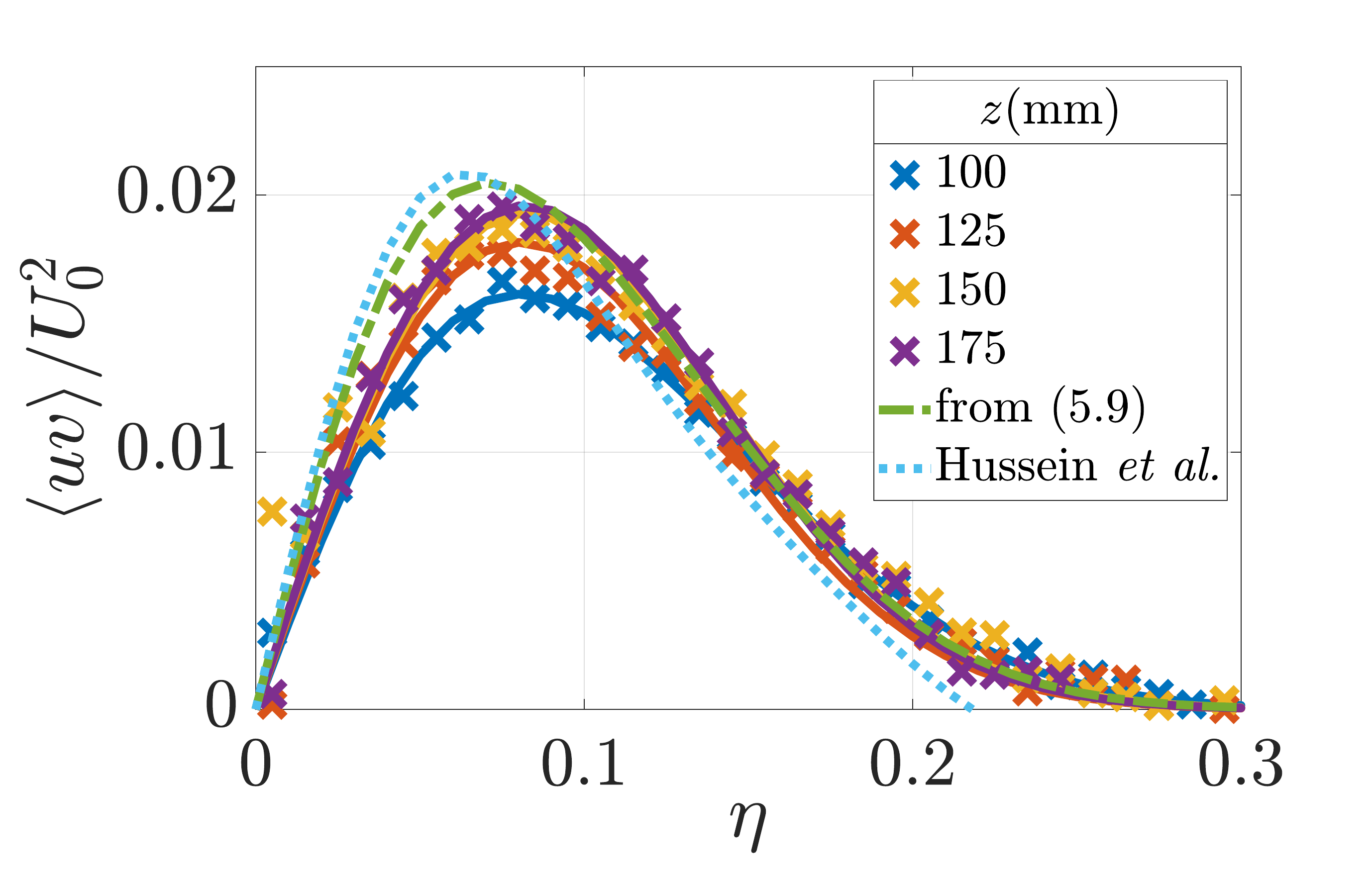}
\caption{(\textit{a})~Self-similar profile $\widehat{\nu}_T(\eta)$ based on relation~\eqref{eq:nu_hat_gen} (solid line: median value, coloured zone limited by dashed lines: 70\% of the measured values). (\textit{b})~Self-similar profile $(\langle uv \rangle / U_0^2)(\eta)$ (crosses and solid lines: experimental points, dashed line: fit based on the relation~\eqref{eq:nu_hat_gen} for $\widehat{\nu}_T$, dotted line: fit from \citet{hussein1994velocity}).\label{fig:nu}}
\end{figure}
Similarly to $\widehat{K}_T$, the turbulent viscosity $\widehat{\nu}_T$ can be estimated from~\eqref{eq:nu_hat_gen} knowing the mean axial velocity profile $f$. Figure~\ref{fig:nu}(\textit{a}) shows the retrieved profile of the turbulent viscosity. As for the turbulent diffusivity, estimates of $\widehat{\nu}_T$ are obtained at various downstream locations $z$. The solid line represents the median value for all $z$ locations, and the coloured zone within the dashed lines comprises 70\% of all measurements. The observed trend, with a relatively constant value near the centreline and an outward decay as $\eta$ increases, is in good qualitative agreement with previous measurements based on the cross-correlation of mean axial and radial velocity fluctuations as presented in \citet{pope2000turbulent}. The centreline value retrieved for $\widehat{\nu}_T$ here, of the order of 0.3, is also in good agreement with the values reported in these previous studies.

Interestingly, going back to the original definition of the turbulent diffusivity based on the cross-correlation of mean axial and radial velocity fluctuations:
\begin{equation}
\widehat{\nu}_T(\eta) = - \dfrac{(\langle uv \rangle / U_0^2)(\eta)}{Sf'(\eta)},
\label{eq:nu_hat_def}
\end{equation}
the previous estimate of $\widehat{\nu}_T(\eta)$ can in turn be used to estimate the self-similar profile of $(\langle uv \rangle / U_0^2)(\eta)$. This is shown in figure~\ref{fig:nu}(\textit{b}), together with the direct measurements of this quantity from the experimental measurements. It can be seen in this figure that, although self-similarity is not perfectly reached yet within the range of accessible streamwise distances, the profile of $\langle uv \rangle / U_0^2$ for the farthest axial distance (corresponding to $z/D \simeq 45$) approaches the profile predicted by~\eqref{eq:nu_hat_gen}. We also show in figure~\ref{fig:nu}(\textit{b}) the profile of $\langle uv \rangle / U_0^2$ fitted by \citet{hussein1994velocity} for their measurements at a streamwise distance of the order of $z/D \simeq 70$, which is found in very good agreement with our prediction (note that their measurements stops at $\eta \simeq 0.2$, hence their proposed fit is not relevant beyond this radial position).

\subsubsection{Determination of $\sigma_T$}
To finish, we propose here an estimate of the radial profile of the turbulent Prandtl number $\sigma_T$. In a situation where $\sigma_T = \nu_T/K_T$ would be uniform (independent of $\eta$), according to relation~\eqref{eq:sigma_uniform} if $f$ is assumed Gaussian (neglecting the aforementioned polynomial correction), then $\Phi$ should also be Gaussian, and the ratio of the half-widths $A_\Phi$ and $A$ for $\Phi$ and $f$, respectively, directly gives an estimate of $\sigma_T$ \citep{corrsin1950further, panchapakesan1993turbulence2, ezzamel2015dynamical}. Using such a Gaussian approximation (light blue fit in figure~\ref{fig:phi_erf}), we obtain $\sigma_T = A_\Phi/A = 0.62$, which is in good agreement with the usual experimental values around 0.7 \citep{pope2000turbulent}.

\begin{figure}
\centerline{\includegraphics[width=0.5\textwidth]{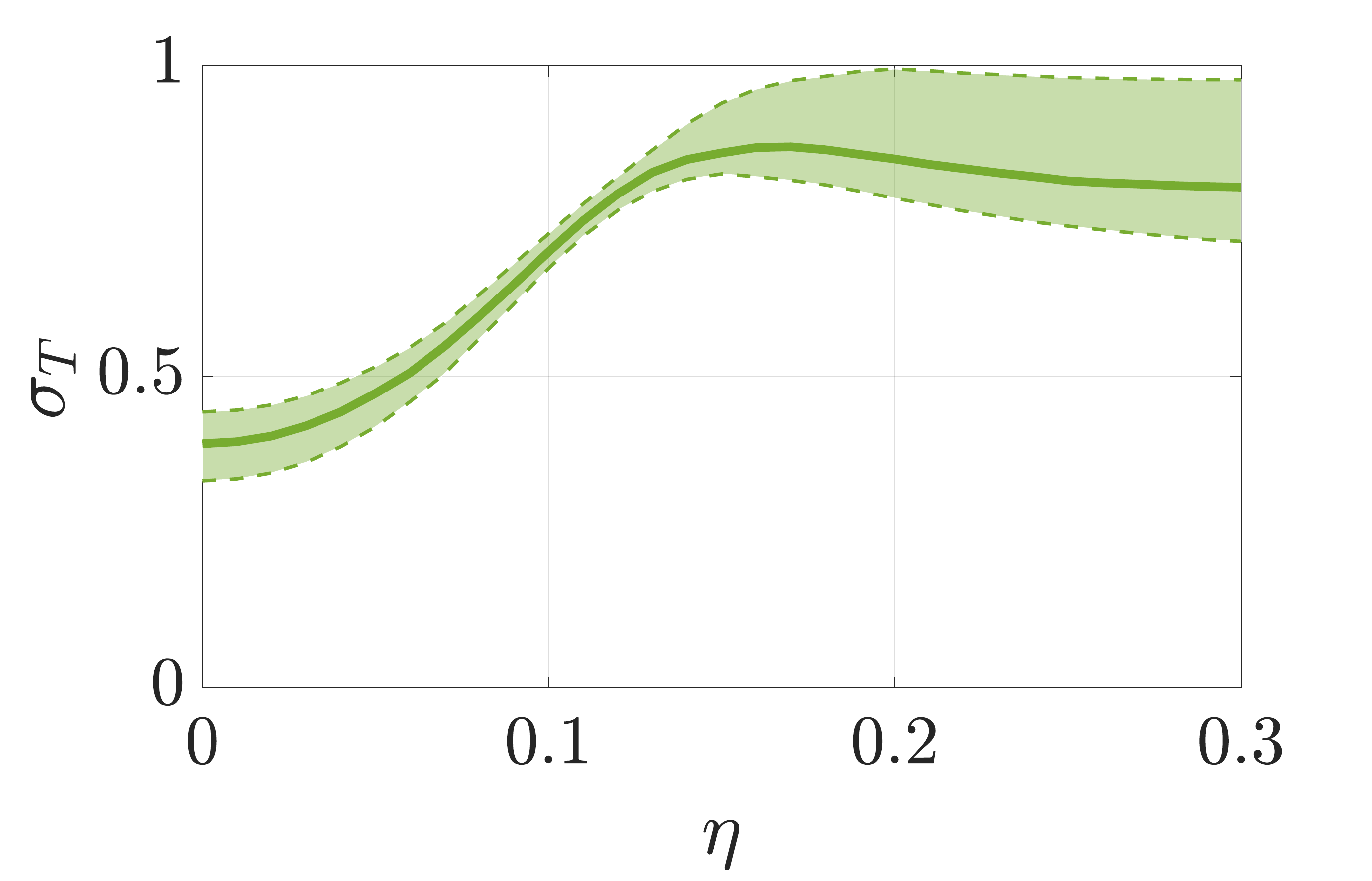}}
\caption{Self-similar profile $\sigma_T(\eta)$ (solid line: median value, coloured zone limited by dashed lines: 70\% of the measured values).\label{fig:sigma}}
\end{figure}
However, the deviation of the concentration profile $\Phi(\eta)$, while $f(\eta)$ is quasi-Gaussian, suggests that $\sigma_T$ may not be considered as uniform. In this case, the profile of $\sigma_T(\eta)$ can be estimated with the generalised relation~\eqref{eq:sigma_nonuniform}, from the simple knowledge of $\Phi$ and $f$. The corresponding profile of $\sigma_T$ is presented in figure~\ref{fig:sigma}. It is actually found to be dependent on $\eta$, increasing between 0.4 near the centreline to an asymptotic value close to 0.8 as larger radial distances are considered, with an average value of the order of 0.6.

The trend of $\sigma_T$ with $\eta$ in previous works is not fully conclusive: \citet{chevray1978intermittency} and \citet{chua1990turbulent} observe a slight increase of $\sigma_T$ with $\eta$, while \citet{chang2002turbulent} report nearly flat then decreasing profile. Direct numerical simulations by \citet{lubbers2001simulation} show a mild increase of $\sigma_T$ with $\eta$ while those by \citet{reeuwijk2016turbulent} show a slight increase then decrease. The lack of consensus regarding the radial dependency of $\sigma_T$ may be related to the sensitivity of $\sigma_T$ determination to experimental and numerical details. The broader-than-Gaussian concentration profile $\Phi$ can for instance be interpreted as a possible effect of the finite size of the particle injection point (at the jet nozzle in the present study), while studies investigating the turbulent diffusion of a passive scalar as temperature \citep{chevray1978intermittency, chua1990turbulent, tong1995passive}, may consider injection points closer to a point source, what seems to lead to Gaussian scalar profiles, hence consistent with a relatively uniform profile of $\sigma_T$. 

With this respect, while all studies are consistent regarding the order of magnitude of $\sigma_T$ and in particular regarding the fact that $\sigma_T < 1$ (i.e. scalar spreads at a slower rate than momentum), the details of any eventual non-uniformity of $\sigma_T$ and whether this is an intrinsic property of the jet or a consequence of experimental/numerical protocols remain to be further clarified. In this perspective, the relations established in the present study, allowing the estimation of turbulent diffusivity, viscosity and Prandtl number from simple measurement of mean concentration and velocity profiles are particularly interesting for future systematic investigations.

\section{Conclusion}\label{sec:conclusion}
Measurements of velocity fields were realised in a free round jet based on Lagrangian tracer trajectories. By using a specific nozzle seeding (where only fluid particles emanating from the nozzle are tagged and not those been entrained into the jet from the surrounding fluid at rest), the self-similar mean velocity profiles were found to differ from those of the global jet (accounting for both, nozzle seeded and entrained fluid particles), in particular for the radial velocity. More precisely, (i)~the nozzle seeded profiles still preserve the self-similar property of the jet, (ii)~the self-similar mean axial velocity profile is not significantly altered by the nozzle seeding compared to the global profile, (iii)~while the self-similar mean radial velocity profile strongly deviates from the usual profile of the global jet.

By revisiting the classical considerations -- connecting global mean axial and radial velocity profiles through the incompressibility of the self-similar jet -- in more general terms of mass conservation, we were able to quantitatively explain the modified self-similar profile. The difference between the global profile and the nozzle seeded profile allows us to specifically identify the contribution associated to the flux of entrained particles to the global mean radial velocity, via a simple entrainment term $\zeta$~\eqref{eq:zeta} solely dependent on the self-similar mean axial velocity profile. This entrained contribution can in turn be interpreted as an effective compressibility for the flow tagged by the nozzle seeded particles. Interestingly, the influence of entrained particles on the mean radial velocity profile is found to be significant up to the core of the jet.

We have then connected this global contribution of entrainment to the classical turbulent advection-diffusion description of the jet. Under the hypothesis of self-similarity, this allowed us to analytically relate turbulent diffusion (of mass and momentum) to the previously identified entrainment term $\zeta$. This results in simple analytical relations~\eqref{eq:K_hat_gen},~\eqref{eq:nu_hat_gen} and~\eqref{eq:sigma_nonuniform} for the turbulent diffusivity $K_T$, the turbulent viscosity $\nu_T$ and the turbulent Prandtl number $\sigma_T$ allowing experimental determination of the non-uniform spatial profiles of these quantities from the simple measurement of the mean scalar (concentration) profile and the mean axial velocity profile. Interestingly, these relations can be used even if the mean concentration and velocity profiles are measured independently as, contrary to classical determinations of turbulent diffusivity based on cross-correlations of velocity and scalar fluctuations, the present relations only require the knowledge of each mean field separately, without requiring to simultaneously measure both fluctuating quantities. Therefore, beyond the fundamental interest of explicitly connecting the entrainment process to turbulent diffusion properties of self-similar jets, these relations can be of real practical interest to experimentally determine the associated diffusion coefficients, including their spatial non-uniformity. In particular, they could help a simple systematic investigation of the non-uniformity of the turbulent Prandtl number for which, while most studies (including the present work) converge to the fact that it is lower than unity (meaning that passive scalar spreads slower than momentum), its eventual spatial dependency remains to be clarified.

Finally, we would like to stress that the approach of the present study, based on a specific inhomogeneous seeding of the flow, intimately connects Lagrangian and Eulerian descriptions of the jet. It shows indeed how tagging particles with a prescribed initial position from which all the Lagrangian trajectories originate affects the corresponding Eulerian fields, which in particular may exhibit an apparent compressibility, even if the global background flow is incompressible. The combination of such a Lagrangian tagging, with first principles such as mass conservation, and in the present case with prescribed properties such as self-similarity, allowed us to gain new insight on the role of entrainment on the mean spreading of the jet, eventually connecting turbulent diffusion properties to the aforementioned effective compressibility.

In future studies, the present inhomogeneous seeding approach could be extended to address higher order turbulent statistics in self-similar jets. For instance, investigating the Eulerian structure functions of the nozzle seeded flow compared to those of the global jet could help disentangling the roles of internal and external intermittency in self-similar jets \citep{gauding2021selfsimilarity}. In a more Lagrangian perspective, having access to longer trajectories (especially through numerical simulations) would enable one to study separately the temporal dynamics of the nozzle seeded particles (from the nozzle to the core of the jet) and of the entrained particles (from outside to inside the jet). It would give access to a Lagrangian understanding of entrainment through the whole space, and not only close to the TNTI. Finally, the approach could also be easily extended to other free shear and/or self-similar flows, such as plane jets, wakes, mixing layers, homogeneous shear flows, grid turbulence, etc.

\backsection[Funding]{B.V., R.V. and M.B. benefit from the financial support of the Project IDEXLYON of the University of Lyon in the framework of the French Programme Investissements d’Avenir (ANR-16-IDEX-0005). T.B., M.G., N.M. and M.B. are supported by French research program ANR-13-BS09-0009 “LTIF”. B.V. and R.B.C. are supported by a US National Science Foundation grant (NSF-GEO-1756259). R.B.C. is also grateful for the support provided through the Fulbright Scholar Program.}

\backsection[Declaration of interests]{The authors report no conflict of interest.}

\appendix
\section{Resolution of the nozzle seeding model}\label{app:A}
We need to solve the continuity equation:
\begin{equation}
\bnabla \bcdot (\langle \varphi \rangle \langle \boldsymbol{U_\varphi} \rangle) = \langle \varphi \rangle \bnabla \bcdot \langle \boldsymbol{U_\varphi} \rangle + \langle \boldsymbol{U_\varphi} \rangle \bcdot \bnabla \langle \varphi \rangle = 0.
\label{eq:A1}
\end{equation}
With the definitions of $U_0(z)$, $\varphi_0(z)$, $f_\varphi(\eta)$, $g_\varphi(\eta)$ and $\Phi(\eta)$ given in the main article, we can show that
\begin{equation}
\langle \varphi \rangle \bnabla \bcdot \langle \boldsymbol{U_\varphi} \rangle = \dfrac{U_0(z) \varphi_0(z)}{r} \Phi(\eta) [(\eta g_\varphi(\eta))' - \eta (\eta f_\varphi(\eta))'],
\label{eq:A2}
\end{equation}
which leads to the usual incompressible solution, and
\begin{equation}
\langle \boldsymbol{U_\varphi} \rangle \bcdot \bnabla \langle \varphi \rangle = \dfrac{U_0(z) \varphi_0(z)}{r} \eta [g_\varphi(\eta) \Phi'(\eta) - f_\varphi(\eta)(\eta \Phi(\eta))'].
\label{eq:A3}
\end{equation}
Thus we get equation~\eqref{eq:continuity_phi_ss} given in the main article:
\begin{equation}
\Phi(\eta) [(\eta g_\varphi(\eta))' - \eta (\eta f_\varphi(\eta))'] + \eta [g_\varphi(\eta) \Phi'(\eta) - f_\varphi(\eta)(\eta \Phi(\eta))'] = 0.
\label{eq:A4}
\end{equation}
Equation~\eqref{eq:A4} can be rewritten as
\begin{equation}
\Phi(\eta) g_\varphi(\eta) + \eta (\Phi(\eta) g_\varphi(\eta))' - \eta^2 (\Phi(\eta) f_\varphi(\eta))' - 2 \eta \Phi(\eta) f_\varphi (\eta) = 0,
\label{eq:A5}
\end{equation}
then
\begin{equation}
(\eta \Phi(\eta) g_\varphi(\eta))' - (\eta^2 \Phi(\eta) f_\varphi(\eta))' = 0.
\label{eq:A6}
\end{equation}
We integrate equation~\eqref{eq:A6} and simplify by $\eta \Phi(\eta)$ (by considering $\eta = 0$, the constant of integration is zero):
\begin{equation}
g_\varphi(\eta) = \eta f_\varphi(\eta).
\label{A7}
\end{equation}

\section{Turbulent quantities from boundary-layer equations}\label{app:B}
In a turbulent free round jet, the mean axial and radial velocity fields, respectively $\langle U \rangle$ and $\langle V \rangle$, are determined with the turbulent boundary-layer equations:
\begin{itemize}
\item the continuity equation:
\begin{equation}
\dfrac{\partial \langle U \rangle}{\partial z} + \dfrac{1}{r} \dfrac{\partial(r \langle V \rangle)}{\partial r} = 0,
\label{eq:B1}
\end{equation}
\item and the Navier-Stokes equation:
\begin{equation}
\langle U \rangle \dfrac{\partial \langle U \rangle}{\partial z} + \langle V \rangle \dfrac{\partial \langle U \rangle}{\partial r} = \dfrac{1}{r} \dfrac{\partial}{\partial r} \left( r \nu_T \dfrac{\partial \langle U \rangle}{\partial r} \right).
\label{eq:B2}
\end{equation}
\end{itemize}
We use the Reynolds decomposition: $U = \langle U \rangle + u$ and $V = \langle V \rangle +v$, and also the gradient closure model $\langle uv \rangle = -\nu_T \dfrac{\partial U}{\partial r}$ (see \citet{pope2000turbulent} or \citet{schlichting2017boundary} for the determination of these equations). Equation~\eqref{eq:B2} is the most simplified writing of the Navier-Stokes equation, and neglects in particular terms in $\langle u^2 \rangle$, $\langle v^2 \rangle$ and $\langle w^2 \rangle$. \citet{hussein1994velocity} experimentally discuss these approximations, and show that it leads to a slight underestimating of $\langle uv \rangle$ and $\nu_T$.

Three quantities are unknown: $\langle U \rangle$, $\langle V \rangle$ and $\nu_T$, with only two equations. Thus we can not solve the system but we can write one quantity as a function of one other, especially we can determine $\nu_T$ as a function of $\langle U \rangle$, or, with the relations introduced in the main article, $\widehat{\nu}_T$ as a function of $f$. We show in the main article than the continuity equation~\eqref{eq:B1} leads to a relation between $f$ and $g$:
\begin{equation}
g(\eta) = \eta f(\eta) - \dfrac{1}{\eta} \int_0^\eta x f(x) \:\mathrm{d}x.
\label{eq:B3}
\end{equation}
Equation~\eqref{eq:B2} can be rewritten with $f$ and $g$:
\begin{equation}
- \eta [f(\eta) (\eta f(\eta))' - g(\eta) f'(\eta)] = S (\eta \widehat{\nu}_T(\eta) f'(\eta))'.
\label{eq:B4}
\end{equation}
We remove $g$ with equation~\eqref{eq:B3}, and the left-hand side term is
\begin{equation}
- \left[ \eta f^2(\eta) + f'(\eta) \int_0^\eta x f(x) \:\mathrm{d}x \right],
\label{B5}
\end{equation}
which can be rewritten as
\begin{equation}
- \left[ f(\eta) \int_0^\eta x f(x) \:\mathrm{d}x \right]'.
\label{eq:B6}
\end{equation}
Thus integration of equation~\eqref{eq:B4} gives
\begin{equation}
\widehat{\nu}_T(\eta) = - \dfrac{1}{S} \dfrac{f(\eta)}{f'(\eta)} \dfrac{1}{\eta} \int_0^\eta x f(x) \:\mathrm{d}x.
\label{eq:B7}
\end{equation}

In the same way, the momentum equation for a conserved passive scalar is
\begin{equation}
\langle U \rangle \dfrac{\partial \langle \varphi \rangle}{\partial z} + \langle V \rangle \dfrac{\partial \langle \varphi \rangle}{\partial r} = \dfrac{1}{r} \dfrac{\partial}{\partial r} \left( r K_T \dfrac{\partial \langle \varphi \rangle}{\partial r} \right).
\label{eq:B8}
\end{equation}
A similar solving leads to
\begin{equation}
\widehat{K}_T(\eta) = - \dfrac{1}{S} \dfrac{\Phi(\eta)}{\Phi'(\eta)} \dfrac{1}{\eta} \int_0^\eta x f(x) \:\mathrm{d}x.
\label{eq:B9}
\end{equation}
Thus $\nu_T$ and $K_T$ are determined with independent calculations, and the general formula of $\sigma_T$ is
\begin{equation}
\sigma_T(\eta) = \dfrac{\nu_T(\eta)}{K_T(\eta)} = \dfrac{\Phi'(\eta)}{\Phi(\eta)} \dfrac{f(\eta)}{f'(\eta)}.
\label{eq:B10}
\end{equation}

\bibliographystyle{jfm}
\bibliography{biblio}

\end{document}